\documentclass[journal]{IEEEtran}
\usepackage{cite}
\usepackage{amsmath,amssymb,amsfonts}
\usepackage[]{footmisc}
\usepackage{xcolor}
\usepackage{graphicx}
\usepackage{textcomp}
\usepackage{psfrag}
\usepackage{epsfig}
\usepackage{subcaption}
\usepackage{float}
\usepackage{xcolor}
\usepackage{tabularx}
\usepackage{booktabs}
\usepackage{algorithm}
\usepackage{algpseudocode}
\usepackage{footmisc}

\DeclareMathOperator*{\argmin}{arg\,min}
\DeclareMathOperator*{\T}{\mathrm{T}}
\DeclareMathOperator*{\F}{\mathrm{F}}
\DeclareMathOperator*{\tr}{\mathrm{tr}}
\DeclareMathOperator*{\rank}{\mathrm{rank}}

\newtheorem{mypro}{Proposition}
\newtheorem{myass}{Assumption}
\newtheorem{myrem}{Remark}

\newtheorem{mylem}{Lemma}
\newtheorem{mythr}{Theorem}

\newtheorem{mycor}{Corollary}

\title{\LARGE A Unified Alternating Optimization Framework for Joint Sensor and Actuator Configuration in LQG Systems}

\begin{document}

\author{Nachuan~Yang,~Yuzhe~Li,~Ling~Shi,~\IEEEmembership{Fellow,~IEEE},~and~Tongwen~Chen,~\IEEEmembership{Fellow,~IEEE}
	\thanks{\rm Nachuan Yang and Tongwen Chen are with the Department of Electrical and Computer Engineering, University of Alberta, Edmonton, AB, Canada T6G 1H9 (email: nachuan1@ualberta.ca, tchen@ualberta.ca).}
	\thanks{\rm Yuzhe Li is with the State Key Laboratory of Synthetical Automation for Process Industries, Northeastern University, Shenyang 110004, China (email: yuzheli@mail.neu.edu.cn).}
    \thanks{\rm Ling Shi is with the Department of Electronic
	and Computer Engineering, with a joint appointment in the Department of Chemical and Biological Engineering, Hong Kong University of Science and
	Technology, Clear Water Bay, Hong Kong SAR (email: eesling@ust.hk).}
}

\maketitle

\begin{abstract}  
This paper fills a gap in the literature by considering a joint sensor and actuator configuration problem under the linear quadratic Gaussian (LQG) performance without assuming a predefined set of candidate components. Different from the existing research, which primarily focuses on selecting or placing sensors and actuators from a fixed group, we consider a more flexible formulation where these components must be designed from scratch, subject to general-form configuration costs and constraints. To address this challenge, we first analytically characterize the gradients of the LQG performance with respect to the sensor and actuator matrices using algebraic Riccati equations. Subsequently, we derive first-order optimality conditions based on the Karush–Kuhn–Tucker (KKT) analysis and develop a unified alternating direction method of multipliers (ADMM)-based alternating optimization framework to address the general-form sensor and actuator configuration problem. Furthermore, we investigate three representative scenarios: sparsity promoting configuration, low-rank promoting configuration, and structure-constrained configuration. For each scenario, we provide in-depth analysis and develop tailored computational schemes. The proposed framework ensures numerical
efficiency and adaptability to various design constraints and
configuration costs, making it well-suited for integration into numerical solvers.
\end{abstract}
% paper that keywords are not normally used for peerreview papers.

\section{Introduction}
The configuration of sensors and actuators is becoming a fundamental issue of modern control engineering, influencing both estimation accuracy and control performance. Traditionally, control research has predominantly assumed sensors and actuators as pre-existing components, focusing primarily on the controller and observer design; whereas in many practical applications, the sensors and actuators themselves must be carefully designed. In applications such as aerospace systems, industrial automation, and large-scale environmental monitoring, the strategic deployment of sensing and actuation devices is crucial for ensuring stability, robustness, and efficiency \cite{stephens2022integrated,meng2023configuration,gao2012allocation}. As systems become increasingly complex and interconnected, traditional heuristic or manually designed configurations are often inadequate, motivating the need for systematic and optimization-based approaches \cite{verdone2010wireless}. Moreover, the constraints on network bandwidth, energy consumption, and real-time adaptability further complicate the configuration of sensors and actuators, necessitating advanced methods that can handle these challenges in a principled manner.

The configuration of sensors and actuators has been investigated in various contexts. Early research focused on sensor placement in static systems for state estimation from noisy data, such as phase measurement
unit (PMU) placement in power grids \cite{nuqui2005phasor}, sensor placement for effective coverage and surveillance in distributed wireless networks \cite{dhillon2003sensor}, and sensor location for health monitoring systems in buildings \cite{meo2005optimal}. These approaches usually fail to work for dynamic structures, which greatly limits their applicability. In the past decade, research on sensor and actuator design issues for dynamical systems has received much attention. In \cite{mo2011sensor}, a multi-step sensor selection strategy was proposed to schedule sensors for Kalman filters. This problem was further studied in \cite{mo2011stochastic} where a stochastic sensor selection algorithm was developed. In \cite{tzoumas2016sensor}, a submodularity-based method was proposed to address the sensor placement for Kalman filters with desired estimation performance. In \cite{tzoumas2015minimal} and \cite{summers2016actuator},  greedy algorithms were developed for actuator selection in networked control systems. In \cite{munz2014sensor}, sensor and actuator placement algorithms were developed for linear time-invariant systems based on $H_2$ and $H_{\infty}$ optimization. However, these existing methods usually consider the sensor placement and actuator placement separately, and a systematic co-design method was not established. 

In recent years, the co-design of sensors and actuators has seen some progress. In \cite{taylor2016allocating}, the authors studied
the placement of actuators and sensors with binary placement and continuous sizing variables. In \cite{tzoumas2020lqg}, the authors addressed the co-design of LQG control and sensing, where the sensing design is selected among a finite set of available sensors. In \cite{huang2021joint}, a branch-and-bound approach was proposed to solve the joint sensor and actuator placement for infinite-horizon LQG control. In \cite{manohar2021optimal}, balanced model reduction and greedy optimization methods were developed to determine a sensor and
actuator selection strategy. However, these methods are only limited to the selection or placement of a predefined group of sensors and actuators, which is not applicable when such a group is not available. This motivates our research.

In this paper, we investigate the joint configuration of sensors and actuators in LQG systems. Unlike existing studies that primarily focus on the placement or selection of sensors and actuators from a predefined set of candidates, we introduce a novel approach where both sensors and actuators are designed from scratch, subject to general-form configuration costs and constraints. The main contributions are as follows:

1) This paper fills the literature gap by considering the joint sensor and actuator configuration in LQG systems, introducing a new formulation without using predefined candidates.

2) The analytical gradients of the LQG performance with respect to the sensor and actuator matrices are derived, followed by the establishment of first-order optimality conditions.

3) A unified ADMM-based alternating optimization framework is developed to solve the joint sensor and actuator configuration problem, ensuring computational tractability while providing adaptability to various design criteria.

4) Three representative scenarios are studied: the sparsity promoting configuration, the low-rank promoting configuration, and the structure-constrained configuration, with tailored analyses and corresponding computation schemes provided.

The remainder of this paper is structured as follows: Section II introduces preliminaries on LQG control and the algebraic Riccati equation, along with our problem formulation. Section III reformulates the problem as a bilevel optimization, derives performance gradients, presents first-order optimality conditions, and develops an ADMM-based alternating optimization scheme. It also discusses three configuration scenarios with tailored analysis and algorithms. Section IV presents a benchmark example and simulations to evaluate performance. Finally,
in Section V, a summary and conclusions are given.

{\it Notations}: The set of real numbers is denoted as $\mathbb{R}$. We use $\mathbb{R}^{m\times n}$ to denote the set of real matrices with $m$ rows and $n$ columns. The notation $[A]_{ij}$ denotes the element located at the $i$-th row and $j$-th column of matrix $A$. The notation $A^{\T}$ denotes the transpose of matrix $A$. The identity matrix and zero matrix are denoted by $I$ and $0$, respectively. The trace of a square matrix $A$ is denoted by $\tr(A)$. The notation $A\succ 0$ (or $A\succeq 0$) means $A$ is positive definite (or semidefinite). The notation $A\succ B$ (or $A\succeq B$) means that $A-B\succ 0$ (or $A-B\succeq 0$). The Frobenius norm and the induced $L_2$ norm of a matrix are denoted by $\|\cdot\|_{\text{F}}$ and $\|\cdot\|_{2}$, respectively. The inner product of two matrices $A$ and $B$ of the same dimensions is denoted as $\langle A,B\rangle = \tr({A^{\T}B})$. The Hadamard product of two matrices $A$ and $B$ is denoted by $A\circ B$ such that $[A\circ B]_{ij}=[A]_{ij}[B]_{ij}$. The notation $(A)_+$ represents the positive part of matrix $A$, i.e., $[(A)_+]_{ij} = \max\{0,[A]_{ij}\}$. The notation $I_S(x)$ represents a binary indicator function such that $I_S(x) = 0$ if $x\in S$ and $I_S(x) = \infty$ if $x\notin S$. The matrices are assumed to have compatible dimensions if not specifically stated.

\section{Preliminaries}
In this section, we present some preliminary knowledge on LQG control and the classical Riccati equation method. Then we formulate the problem of interest in this paper.
\subsection{Linear Quadratic Gaussian Control}
Consider a linear time-invariant (LTI) system described by the following ordinary differential equation:
\begin{equation}\label{sys}
    \begin{split}
        &\dot{x}(t) = Ax(t)+Bu(t)+w(t),\\
        &y(t) = Cx(t) + v(t),
    \end{split}
\end{equation}
where $x(t)\in\mathbb{R}^n$ is the system state, $u(t)\in\mathbb{R}^m$ is the control input, and $y(t)\in\mathbb{R}^q$ denotes the sensor measurement. $A,B,C$ are system matrices. Here, $w(t)\in\mathbb{R}^n$ and $v(t)\in\mathbb{R}^q$ denote independent white Gaussian noises such that 
\begin{align*}
    \mathbb{E}[w(t)w(t)^{\mathrm{T}}] = \Pi_w,~~~\mathbb{E}[v(t)v(t)^{\mathrm{T}}] = \Pi_v,
\end{align*}
where $\Pi_w,\Pi_v\succ 0$ denote the corresponding noise covariance matrices. Since only partial state information is available, the following Luenberger observer is used for state estimation:
\begin{equation}
    \begin{split}
        &\dot{\hat{x}}(t) = A\hat{x}(t)+Bu(t)+L(\hat{y}(t) - y(t))\\
        &\hat{y}(t) = C\hat{x}(t)
    \end{split}
\end{equation}
where $\hat{x}(t)\in\mathbb{R}^n$ denotes the state estimate and $L\in\mathbb{R}^{n\times q}$ is the observer gain to be determined. Then the dynamic output-feedback controller can be designed as follows:
\begin{align}
    u(t) = K\hat{x}(t)
\end{align}
where $K\in \mathbb{R}^{m\times n}$ is the feedback gain matrix to be determined. The linear quadratic Gaussian control problem aims at optimizing the following performance index:
\begin{align}\label{lqg}
     J_{\text{LQG}} := \lim_{T\rightarrow \infty} \frac{1}{T}\mathbb{E}[\int_{t=0}^{T}x(t)^{\mathrm{T}}Qx(t) + u(t)^{\mathrm{T}}Ru(t)~dt]
\end{align}
where $Q,R\succ 0$ denote the relative weight matrices for the system state and control input, respectively. Throughout this paper, we assume that $\Pi_w$ and $Q$ are strictly positive definite to facilitate theoretical analysis, while noting that the proposed approach can be extended to the case where $\Pi_w,Q \succeq 0$ without affecting the main contributions of this work.

\subsection{Algebraic Riccati Equation Method}
It is known that under certain standard assumptions, the optimal solution to the infinite-horizon LQG control problem in \eqref{lqg} can be characterized via a pair of continuous-time algebraic Riccati equations (CAREs). To establish the existence and uniqueness of the optimal solution, we impose the following assumption that is widely adopted in the LQG control \cite{anderson2007optimal}.
\vspace{3pt}

\begin{myass}
    $(A,B)$ is stabilizable; $(A,C)$ is detectable.
\end{myass}
\vspace{3pt}

Under Assumption 1, the existence of unique positive definite solutions to the CAREs is guaranteed. The optimal gains can be obtained by solving the following dual CAREs:
\begin{align}
   &AX+XA^{\mathrm{T}}-XC^{\mathrm{T}}\Pi_v^{-1}CX+\Pi_w=0, \label{care1}\\
    &A^{\mathrm{T}}P+PA-PBR^{-1}B^{\mathrm{T}}P +Q = 0.\label{care2}
\end{align}
where $X,P\succ 0$ denote the corresponding solution matrices. Then the optimal observer gain (also known as the Kalman gain) and controller gain can be obtained as
\begin{align}\label{optlk}
 L^* = - XC^{\mathrm{T}}\Pi_v^{-1} ~~\text{ and } ~~K^* = -R^{-1}B^{\mathrm{T}}P.
\end{align}
The corresponding LQG performance is given by
\begin{align}\label{lqg1}
   J_{\text{LQG}} = \tr(XQ)+\tr(PXC^{\mathrm{T}}\Pi_v^{-1}CX),
\end{align}
or equivalently 
\begin{align}\label{lqg2}
   J_{\text{LQG}} = \tr(P\Pi_{w})+\tr(XPBR^{-1}B^{\mathrm{T}}P),
\end{align}
which are derived from the controller and observer perspectives, respectively \cite{arelhi1997lqg}. A fundamental property of LQG control is that the CAREs in \eqref{care1} and \eqref{care2} can be solved independently, leading to separated design of optimal observer and controller gains, as shown in \eqref{optlk}. This decoupling property is known as the separation principle in control theory.
\subsection{Sensor and Actuator Configuration Problem}
In the conventional setup of LQG control, the actuator and sensor matrices are typically given, and the focus is on designing the controller. However, in many practical applications, the configuration of sensors and actuators is not predefined, but must be strategically designed by users to meet specific requirements and optimize system performance \cite{stephens2022integrated}.

In this paper, we are interested in the joint configuration of sensors and actuators in an LQG setup. More specifically, we consider a general formulation by incorporating both the configuration costs and constraints as follows:
\begin{align}
    J_{\text{SAC}} = \Phi(B) + \Psi(C),~B \in \Omega_B,~C\in \Omega_C,
\end{align}
where $J_{\text{SAC}}$ denotes the total cost of the sensor and actuator configuration. Here, $\Phi(\cdot)$ and $\Psi(\cdot)$ are convex functions used to evaluate the configuration costs of actuators and sensors. $\Omega_B$ and $\Omega_C$ are convex sets representing constraints such as hardware limitations. Throughout this paper, we assume that noises $w(t),v(t)$ are independent of the configuration, or its influence is negligible.
To guarantee the existence of LQG controllers, we impose the following natural constraints:
\begin{align*}
    &B \in \mathcal{S} := \{B\in\mathbb{R}^{n\times m}\mid (A,B)\text{ is stabilizable}\},\\
    &C \in \mathcal{D} := \{C\in\mathbb{R}^{q\times n}\mid (A,C)\text{ is detectable}\},
\end{align*}
which represent the stabilizability and detectability domains of system \eqref{sys}. Then, the optimal observer and controller gains are uniquely determined by CAREs, if $B \in \mathcal{S}$ and $C \in \mathcal{D}$ are fixed. As a result, the LQG performance $J_{\text{LQG}}$ is well defined, and the overall cost can be described by
\begin{align}
    J(B,C) := J_{\text{LQG}} + \gamma J_{\text{SAC}},
\end{align}
where $\gamma>0$ denotes the relative weight for configuration cost. In what follows, we formulate the sensor and actuator configuration problem to be investigated in this paper.
\vspace{3pt}

\textbf{Problem of Interest:} Regarding a linear Gaussian process as described in system \eqref{sys}, determine an actuator matrix $B \in \Omega_B$ and a sensor matrix $C \in \Omega_C$ such that the overall control system is stabilizable and detectable, while minimizing the weighted LQG performance and configuration cost, i.e.,
\begin{equation}\tag{P0}\label{p0}
    \begin{split}
        \min_{B\in \mathcal{S},C\in \mathcal{D}}J(B,C)
        ~\text{ subject to } B \in \Omega_B,~C\in \Omega_C,
    \end{split}
\end{equation}
where $J(B,C) := J_{\text{LQG}} + \gamma J_{\text{SAC}}$ with 
\begin{align*}
J_{\text{SAC}} &= \Phi(B) + \Psi(C),\\
    J_{\text{LQG}} &= \tr(XQ)+\tr(PXC^{\mathrm{T}}\Pi_v^{-1}CX)\\
    &=\tr(P\Pi_{w})+\tr(XPBR^{-1}B^{\mathrm{T}}P)
\end{align*}
and matrices $X\succeq 0, P\succeq 0$ are uniquely determined by the continuous-time algebraic Riccati equations in \eqref{care1} and \eqref{care2}.
\vspace{3pt}

\begin{myrem}
The above formulation outlines a general joint sensor and actuator configuration problem for LQG systems, where the configuration costs and constraints are unspecified. The challenges mainly arise from three aspects: (1) the intricate relationship between LQG performance and the actuator/sensor matrices, which lacks an explicit analytical form; (2) the nonconvexity and potential nonsmoothness introduced by both the LQG performance and the configuration cost functions; and (3) the coupling between sensor and actuator matrices, which further complicates the co-design process.
\end{myrem}

\section{Main Results}
In this section, we first investigate Problem (P0) within a bi-level optimization framework and propose an ADMM-based alternating optimization algorithm to tackle the joint sensor and actuator configuration problem. Subsequently, we present and discuss three representative scenarios: sparsity promoting configuration (SPC), low-rank promoting configuration (LPC), and structure-constrained configuration (SCC), each characterized by distinct configuration costs and constraints. The corresponding analyses and tailored algorithms for these scenarios are detailed in the subsequent subsections.

\subsection{Optimization-based Sensor and Actuator Configuration}

\noindent To mitigate the strong coupling between the sensor and actuator matrices $B$ and $C$ in the original problem, we reformulate and analyze the problem within the framework of bi-level optimization. This approach enables the matrix variables to be separated into two distinct layers, thereby facilitating a more manageable optimization process while preserving the general optimality of the original problem. Specifically, Problem \eqref{p0} can be transformed into two equivalent bi-level optimization problems, each of which provides a different perspective on the sensor and actuator configuration task:
\begin{equation}\label{p1} \tag{P1}
    \begin{aligned}
        \min_{B \in \mathcal{S}} \quad & J_{\text{LQG}}(B, C^{\dagger}) + \gamma\Phi(B) \\
        \text{subject to} \quad & B \in \Omega_B, \\
        & C^\dagger \in \argmin_{C \in \Omega_C} \, \{J_{\text{LQG}}(B, C) + \gamma\Psi(C)\}.
    \end{aligned}
\end{equation}
\noindent Or equivalently,
\begin{equation}\label{p2} \tag{P2}
    \begin{aligned}
        \min_{C \in \mathcal{D}} \quad & J_{\text{LQG}}(B^\dagger, C) + \gamma\Psi(C) \\
        \text{subject to} \quad & C \in \Omega_C, \\
        & B^\dagger \in \argmin_{B \in \Omega_B } \{J_{\text{LQG}}(B, C) + \gamma\Phi(B)\}.
    \end{aligned}
\end{equation}
\noindent In these formulations, Problem \eqref{p1} optimizes $B$ in the upper level and $C$ in the lower level, while Problem \eqref{p2} reverses this order. Here, we use the notation $\dagger$ to denote the solutions to sub-problems. The bi-level reformulations highlight the interdependence between the optimization of variables $B$ and $C$. Before proceeding with the development of the algorithm to solve Problem \eqref{p0}, we first explore the analytical properties of the LQG performance from an optimization perspective.
\vspace{3pt}

\begin{mythr}
    Given a constant sensor matrix $C \in \mathcal{D}$, for any actuator matrix $B \in \mathcal{S}$, the gradient of the LQG performance $J_{\text{LQG}}$ with respect to matrix $B$ is given by
    \begin{align}
        \frac{\partial J_{\text{LQG}}}{\partial B} = - 2P(G_1+G_2)PBR^{-1}.
    \end{align}
    Here, $G_1$ and $G_2$ are the unique symmetric solutions to the following Lyapunov matrix equations
    \begin{align*}
        &G_1(A+BK^*)^{\T} + (A+BK^*)G_1 + \Pi_{w} = 0,\\
        &G_2(A+BK^*)^{\T} + (A+BK^*)G_2 + XA^{\T}+AX = 0,
    \end{align*}
    with $K^* = -R^{-1}B^{\T}P$ denoting the optimal controller gain.
\end{mythr}
\vspace{3pt}
Proof. Consider the increment of the algebraic Riccati equation in \eqref{care2} with $P = P+\Delta P$ and $B = B+\Delta B$, we have 
\begin{align*}
    A^{\T}\Delta P &+ \Delta P A^{\T} - \Delta PBR^{-1}B^{\T}P - PBR^{-1}B^{\T}\Delta P\\
    &~~~~~~~- P\Delta BR^{-1}B^{\T} P - P BR^{-1}\Delta B^{\T} P =0.
\end{align*}
By reorganizing the above equation, we further obtain
\begin{equation}\label{nc1}
    \begin{split}
        (A- BR^{-1}B^{\T}&P)^{\T}\Delta P + \Delta P (A- BR^{-1}B^{\T}P) \\
    &= 
P\Delta BR^{-1}B^{\T} P + P BR^{-1}\Delta B^{\T} P.
    \end{split}
\end{equation}
Post-multiplying matrix $G_1$ to the above equation and taking traces on both sides, we have the following equation:
\begin{align*}
    \tr(\Delta P \Pi_w) &= -\tr(P\Delta BR^{-1}B^{\T} PG_1 +P BR^{-1}\Delta B^{\T} P G_1)\\
    &= - 2\tr(R^{-1}B^{\T}PG_1P\Delta B)\\
    &= \langle - 2PG_1PBR^{-1}, \Delta B\rangle,
\end{align*}
which follows from $A- BR^{-1}B^{\T}P = A + BK^*$. Therefore based on the increment analysis, we can obtain the gradient of $\tr(P\Pi_w)$ with respect to matrix $B$ as follows:
\begin{align}
    \frac{\partial \tr(P\Pi_w)}{\partial B} = - 2PG_1PBR^{-1}.
\end{align}
Then we analyze the gradient of the second term in the LQG performance. By Eqn \eqref{care2}, we always have 
\begin{align}
    PBR^{-1}B^{\mathrm{T}}P = A^{\mathrm{T}}P+PA +Q.
\end{align}
Hence we can equivalently consider $\tr(X(A^{\mathrm{T}}P+PA +Q))$ instead. Post-multiplying Eqn \eqref{nc1} with matrix $G_2$ and taking traces on both sides, we obtain the following equation:
\begin{align*}
    \tr(\Delta P (XA^{\T} + AX)) &= - 2\tr(R^{-1}B^{\T}PG_2P\Delta B).
\end{align*}
By reorganizing the above equations, we have that
\begin{align*}
    \Delta &\tr(X(A^{\mathrm{T}}P+PA +Q)) = \tr(X(A^{\mathrm{T}}\Delta P+\Delta PA))\\
    &= \tr(\Delta P (XA^{\T} + AX)) = \langle - 2PG_2PBR^{-1}, \Delta B\rangle,
\end{align*}
which further implies that 
\begin{align}
    \frac{\partial \tr(XPBR^{-1}B^{\mathrm{T}}P)}{\partial B} = - 2PG_2PBR^{-1}.
\end{align}
Therefore we conclude that the gradient of the LQG performance $J_{\text{LQG}}$ with respect to matrix $B$ is given by
\begin{align*}
    \frac{\partial J_{\text{LQG}}}{\partial B} &= \frac{\partial \tr(P\Pi_w)}{\partial B}+\frac{\partial \tr(XPBR^{-1}B^{\mathrm{T}}P)}{\partial B}\\
    &= - 2P(G_1+G_2)PBR^{-1}.
\end{align*}
The proof is completed. $\hfill\square$
\vspace{3pt}

\begin{mythr}
    Given a constant actuator matrix $B \in \mathcal{D}$, for any sensor matrix $C \in \mathcal{S}$, the gradient of the LQG performance $J_{\text{LQG}}$ with respect to matrix $C$ is given by
    \begin{align}
        \frac{\partial J_{\text{LQG}}}{\partial C} = -2\Pi_v^{-1}CX(H_1+H_2)X.
    \end{align}
    Here, $H_1$ and $H_2$ are the unique symmetric 
 solutions to the following Lyapunov matrix equations
    \begin{align*}
        &H_1(A+L^*C) + (A+L^*C)^{\T}H_1 + Q = 0,\\
        &H_2(A+L^*C) + (A+L^*C)^{\T}H_2 + PA+A^{\T}P = 0,
    \end{align*}
    with $L^* = - XC^{\mathrm{T}}\Pi_v^{-1}$ denoting the optimal observer gain.
\end{mythr}
\vspace{3pt}
Proof. Consider the increment of the algebraic Riccati equation in \eqref{care1} with $X = X+\Delta X$ and $C = C+\Delta C$, we have
\begin{align*}
    A\Delta X + \Delta X& A^{\T} -\Delta XC^{\T}\Pi_v^{-1}CX - XC^{\T}\Pi_v^{-1}C\Delta X\\
    &-X\Delta C^{\T}\Pi_v^{-1}CX- XC^{\T}\Pi_v^{-1}\Delta CX = 0.
\end{align*}
By reorganizing the above equations, we obtain that
\begin{equation}\label{nc2}
\begin{split}
      (A-XC^{\T}&\Pi_v^{-1}C)\Delta X + \Delta X(A- XC^{\T}\Pi_v^{-1}C)^{\T}\\
    &= X\Delta C^{\T}\Pi_v^{-1}CX+XC^{\T}\Pi_v^{-1}\Delta CX.
\end{split}
\end{equation}
Post-multiplying matrix $H_1$ to the above equation and taking traces on both sides, we further obtain that
\begin{align*}
   \Delta \tr(XQ) &= \tr(\Delta X Q) \\
   &= -\tr(X\Delta C^{\T}\Pi_v^{-1}CXH_1 + XC^{\T}\Pi_v^{-1}\Delta CXH_1)\\
    &= - 2\tr(XH_1XC^{\T}\Pi_v^{-1} \Delta C)\\
    &= \langle -2\Pi_v^{-1}CXH_1X, \Delta C\rangle,
\end{align*}
which further implies that 
\begin{align}
    \frac{\partial \tr(XQ) }{\partial C} = -2\Pi_v^{-1}CXH_1X.
\end{align}
Then we analyze the gradient of the second term in the LQG performance. By Eqn \eqref{care1}, we always have 
\begin{align}
    XC^{\mathrm{T}}\Pi_v^{-1}CX = AX+XA^{\mathrm{T}}+\Pi_w.
\end{align}
Similarly, post-multiplying Eqn \eqref{nc2} with $H_2$ and taking traces on both sides, we obtain the following equation:
\begin{align*}
   \Delta \tr(PXC^{\mathrm{T}}\Pi_v^{-1}CX) &= \Delta \tr(P(AX+XA^{\mathrm{T}}+\Pi_w))\\
   &= \tr(\Delta X (PA + A^{\T}P)) \\
    &= \langle -2\Pi_v^{-1}CXH_2X, \Delta C\rangle,
\end{align*}
which further implies that
\begin{align}
    \frac{\partial \tr(PXC^{\mathrm{T}}\Pi_v^{-1}CX) }{\partial C} = -2\Pi_v^{-1}CXH_2X.
\end{align}
Therefore we conclude that the gradient of the LQG performance $J_{\text{LQG}}$ with respect to matrix $C$ is given by
\begin{align*}
    \frac{\partial J_{\text{LQG}}}{\partial C} &= \frac{\partial \tr(XQ)}{\partial C}+\frac{\tr(PXC^{\mathrm{T}}\Pi_v^{-1}CX)}{\partial C}\\
    &= -2\Pi_v^{-1}CX(H_1+H_2)X.
\end{align*}
Thus the proof is completed. $\hfill\square$
\vspace{3pt}

The above two theorems present the methods for computing the gradients of the LQG performance with respect to the sensor and actuator matrices using Lyapunov matrix equations. With the characterization of gradients, we can not only evaluate the solution optimality of the configuration problem, but also develop efficient numerical algorithms to handle the constraints and optimize the objective function.
\vspace{3pt}

\begin{mythr}
    If a matrix pair $(B,C)$ is an optimal solution to Problem \eqref{p0}, it must satisfy the following conditions:
    \begin{align*}
        &0 \in \gamma\frac{\partial \Phi(B)}{\partial B} + \mathcal{N}_{\Omega_B}(B) - 2 P(G_1+G_2)PBR^{-1},\\
        &0\in \gamma\frac{\partial \Psi(C)}{\partial C} + \mathcal{N}_{\Omega_C}(C)-2\Pi_v^{-1}CX(H_1+H_2)X,
    \end{align*}
    where $\mathcal{N}_{\Omega_B}(B)$ and $\mathcal{N}_{\Omega_C}(C)$ are normal cones defined by
    \begin{align*}
        &\mathcal{N}_{\Omega_B}(B) = \{\Lambda_B \mid \forall B' \in \Omega_{B},~\tr(\Lambda_B^{\T} (B' - B)) \le 0\},\\
        &\mathcal{N}_{\Omega_C}(C) = \{\Lambda_C \mid \forall C' \in \Omega_{C},~\tr(\Lambda_C^{\T} (C' - C)) \le 0\}.
    \end{align*}
    More specifically, if the constraint sets $\Omega_B$ and $\Omega_C$ can be explicitly described in the following form:
    \begin{align*}
        &\Omega_B = \{B\mid g_B^i(B) = 0,~h_B^j(B) \le 0\},\\
        &\Omega_C = \{C\mid g_C^i(C) = 0,~h_C^j(C) \le 0\},
    \end{align*}
    then the following conditions hold:
    \begin{align*}
        &0\in \gamma\frac{\partial \Phi(B)}{\partial B}  - 2 P(G_1+G_2)PBR^{-1} \\
        &~~~~~~~~~~~~~~~~~~~+ \sum_{i} \lambda_B^i \frac{\partial g_B^i(B)}{\partial B} + \sum_{j} \mu_B^j \frac{\partial h_B^j(B)}{\partial B},\\
         &0\in \gamma\frac{\partial \Psi(C)}{\partial C} -2\Pi_v^{-1}CX(H_1+H_2)X \\
        &~~~~~~~~~~~~~~~~~~~+ \sum_{i} \lambda_C^i \frac{\partial g_C^i(C)}{\partial C} + \sum_{j} \mu_C^j \frac{\partial h_C^j(C)}{\partial C},\\
        &\mu_B^j \ge 0,~\mu_C^j\ge 0,~\sum_{j} \mu_B^jh_B^j(B) =0,~\sum_j \mu_C^jh_C^j(C)=0,
    \end{align*}
    if $(B,C)$ is an optimal solution to Problem \eqref{p0}.
\end{mythr}
\vspace{3pt}
Proof. 
Let $(B,C)$ be an optimal solution to Problem \eqref{p0}. According to the variational inequality theory \cite{boyd2004convex}, for any constrained optimization problem $\min_{x}f(x)~\text{subject to}~x\in\Omega$, if $x$ is an optimal solution, then
\begin{equation}
    0 \in \partial f(x) + \mathcal{N}_{\Omega}(x),
\end{equation}
where $\partial f(x)$ is the subdifferential of the objective function and $\mathcal{N}_{\Omega}(x)$ is the normal cone to the constraint set at $x$.
Applying this principle to our problem, let $Y:=\begin{bmatrix}
    B^{\T} & C^{\T}
\end{bmatrix}^{\T}$ denote the overall variable, we have the following condition:
\begin{align}
    0 &\in \frac{\partial J_{\text{LQG}}(Y)}{\partial Y} +\gamma \frac{\partial \Phi(B)}{\partial Y} + \gamma\frac{\partial \Psi(C)}{\partial Y} + \mathcal{N}_{\Omega_Y}(Y) ,
\end{align}
where $\mathcal{N}_{\Omega_Y}(Y)$ is the normal cone of set $\Omega_Y := \Omega_B \times \Omega_C$. By matrix calculus and Theorems 1 and 2, we have that
\begin{align}
    \frac{\partial J_{\text{LQG}}(Y)}{\partial Y} = \begin{bmatrix}
        - 2 P(G_1+G_2)PBR^{-1}\\
        -2\Pi_v^{-1}CX(H_1+H_2)X
    \end{bmatrix},
\end{align}
and 
\begin{align*}
    \frac{\partial \Phi(B)}{\partial Y} = \begin{bmatrix}
        \partial \Phi(B)/\partial B\\
        0
    \end{bmatrix},~~\frac{\partial \Psi(C)}{\partial Y} = \begin{bmatrix}
        0\\
        \partial \Psi(C)/\partial C
    \end{bmatrix}.
\end{align*}
Further notice that the set $\mathcal{N}_{\Omega_Y}$ can be given by
\begin{align*}
    \mathcal{N}_{\Omega_Y} = \{\Lambda_Y:= \begin{bmatrix}
        \Lambda_B\\
        \Lambda_C
    \end{bmatrix}\mid  \forall Y'\in\Omega_Y,~\tr(\Lambda_Y^{\T} (Y'-Y))\le 0\}.
\end{align*}
Therefore we have $\Lambda_Y\in \mathcal{N}_{\Omega_Y}$ if and only if
\begin{align}\label{gs}
    \tr(\Lambda_B^{\T}(B'-B)) + \tr(\Lambda_C^{\T}(C'-C))\le 0
\end{align}
for any $B'\in \Omega_B$ and $C'\in\Omega_C$. Notice that if we take $B'=B$ and $C'=C$ respectively, we obtain that 
\begin{align}
   &\forall B'\in \Omega_B, ~\tr(\Lambda_B^{\T}(B'-B))\le 0, \\
    &\forall C'\in\Omega_C,~ \tr(\Lambda_C^{\T}(C'-C))\le 0.
\end{align}
Conversely, if the above two conditions hold, then \eqref{gs} holds and thus the equivalence holds. Therefore, we can obtain the general-form optimality conditions stated in Theorem 3. For the specific cases where $\Omega_B$ and $\Omega_C$ can be described using equality and inequality constraints, we note that
\begin{align*}
    &\mathcal{N}_{\Omega_B}(B) = \left\{\sum_i \lambda_B^i \frac{\partial g_B^i(B)}{\partial B} + \sum_j \mu_B^j \frac{\partial h_B^j(B)}{\partial B} \mid \mu_B^j \geq 0\right\},\\
    &\mathcal{N}_{\Omega_C}(C) = \left\{\sum_i \lambda_C^i \frac{\partial g_C^i(C)}{\partial C} + \sum_j \mu_C^j \frac{\partial h_C^j(C)}{\partial C} \mid \mu_C^j \geq 0\right\}.
\end{align*}
By substituting the above expressions into the general conditions and including the complementary slackness conditions, we can finally obtain the specific optimality conditions stated in Theorem 3. The proof is completed. $\hfill \square$
\vspace{3pt}

\begin{myrem}
   It is worth noticing that the optimality conditions established in Theorem 3 are necessary but not sufficient. Consequently, a solution satisfying these conditions may not be globally optimal but rather represents a stationary point, indicating local optimality. In the context of nonconvex optimization problems, finding a stationary solution is regarded as a primary objective in numerical computation. While Theorem 3 does not directly yield a computational approach, it provides a straightforward way for us to verify stationary solutions.
\end{myrem}
\vspace{3pt}

\begin{mypro}
    The LQG performance $J_{\text{LQG}}$ tends to infinity as the actuator matrix $B\in\mathcal{S}$ approaches the boundary of $\mathcal{S}$ or the sensor matrix $C\in \mathcal{D}$ approaches the boundary of $\mathcal{D}$.
\end{mypro}
\vspace{3pt}
Proof. Intuitively, this follows from the fact that as the actuator matrix $B$ approaches the boundary of $\mathcal{S}$ or the sensor matrix $C$ approaches the boundary of $\mathcal{D}$, the closed-loop matrix $A + BK^*$ (or $A + LC^*$) tends to become unstable, causing the LQG performance to diverge to infinity. 

Here, we provide a detailed mathematical proof within the context of our optimization setup. Let us consider a sequence $\{B_i\}_{i=1}^{\infty} \subset \mathcal{S}$ such that $\lim_{i \rightarrow \infty} B_i = B \in \partial \mathcal{S}$ where $\partial \mathcal{S}$ denotes the boundary of $\mathcal{S}$. Since the spectral abscissa $\sigma(\cdot)$ is a continuous function, it follows that
\begin{align}
    \sigma(A+B_iK_i^*) \rightarrow \sigma (A+BK^*)=0~~\text{as }i\rightarrow\infty,
\end{align}
where $K_i^* = -R^{-1}B_i^{\T}P$ denotes the optimal gain matrix corresponding to matrix $B_i$. Therefore for any $\epsilon >0$, there exists a sufficiently large integer $N$ such that 
\begin{align}
    \forall i>N,~|\sigma(A+B_iK_i^*)|<\epsilon.
\end{align}
Notice that for each actuator matrix $B_i$, the algebraic Riccati equation in \eqref{care2} can be equivalently rewritten as
\begin{align*}
    (A+B_iK_i^*)^{\T}P_i+P_i(A_i+B_iK_i^*) +K_i^*RK_i^*+Q = 0.
\end{align*}
Based on the well-known lower bounds for Lyapunov equations given in \cite{yasuda1979upper}, we can obtain the following result:
\begin{align}
   \tr(P_i)\ge \lambda_{{\text{max}}}(P_i) \ge \frac{\lambda_{\text{min}}(Q)}{-2\sigma(A+B_iK_i^*)}.
\end{align}
By Eqn \eqref{lqg2}, we further have that
\begin{align}
    J_{\text{LQG}} \ge \lambda_{\text{min}}(\Pi_w)\tr(P_i) \ge \frac{\lambda_{\text{min}}(\Pi_w)\lambda_{\text{min}}(Q)}{2\epsilon},
\end{align}
thus $J_{\text{LQG}} \rightarrow \infty$ as $\epsilon \rightarrow 0$, i.e., the LQG performance tends to infinity as $B$ approaches the boundary of $\mathcal{S}$. The proof for the sensor matrix $C$ is similar and is omitted here. 
$\hfill \square$
\vspace{3pt}

\begin{myrem}
  The above proposition paves the way for using gradient-based methods to ensure the stabilizability and detectability of the sensor and actuator matrices. More specifically, as the descent algorithm approaches the boundary of $\mathcal{S}$ or $\mathcal{D}$, the rapidly increasing objective function value generates steep gradients, exhibiting the so-called coercive property \cite{nocedal1999numerical}. This widely used property in optimization theory ensures that the solution is naturally directed back toward the feasible region, preventing the algorithm from generating unstabilizable or undetectable solutions. By leveraging this coercive feature, the algorithm effectively stays within the constraints, ensuring the stabilizability and detectability conditions are satisfied.
\end{myrem}
\vspace{3pt}

Based on the bilevel formulations \eqref{p1} and \eqref{p2}, we propose an alternating optimization scheme to find the optimal sensor and actuator matrices, which is a common method use in bilevel optimization problems \cite{bard2013practical}. The core idea of this method is that, leveraging the bi-level structure, we can iteratively optimize one matrix at the upper level while fixing the other at the lower level, leading to a sequence of simpler subproblems that are easier to solve. This two-layer optimization approach naturally decouples the problem, enabling the solution to be progressively refined by alternating between the sensor optimization and actuator optimization. This alternating procedure can be expressed as
\begin{align}
B_{h+1} &= \argmin_{B\in\mathcal{S}} J_{\text{LQG}}(B, C_h) + \gamma\Phi(B)+I_{\Omega_B}(B),\label{sub1} \\ 
C_{h+1} &= \argmin_{C\in\mathcal{D}} J_{\text{LQG}}(B_{h+1}, C) + \gamma\Psi(C)+I_{\Omega_C}(C),
\label{sub2}
\end{align}
where $B_0$ and $C_0$ are initialized such that the control system is both stabilizable and detectable. In what follows, we will demonstrate how to solve each subproblem in this alternating optimization scheme and evaluate its numerical optimality.
\begin{algorithm}[H]
\caption{Gradient Descent with Backtracking Line Search}
\label{alg:pgd_backtracking}
\begin{algorithmic}[1]
    \State Set Armijo parameter $\alpha \in (0,1)$.
    \State Set backtracking parameter $\beta \in (0,1)$.
   \State Initialize $B_0 \in \mathcal{S}$ and iterator $i \gets 0$.
\While{not convergent}
    \State Compute the following gradient
    \begin{align*}
        \nabla f(B_i) = \frac{\partial \mathcal{L}_\rho(B, M^k, L_B^k)}{\partial B}\mid_{B = B_i}.
    \end{align*}
    % \State Compute projection: $\tilde{B}_i \gets \Pi_{\Omega_B}(B_i - \nabla f(B_i))$.
    % \State Set descent direction: $D_i \gets \tilde{B}_i - B_i$.
    \State Initialize step size $t_i \gets 1$.
    \While{$$f(B_i - t_i \nabla f(B_i)) > f(B_i) - \alpha t_i \tr(\nabla f(B_i)^T \nabla f(B_i))$$}
        \State $t_i \gets \beta t_i$.
    \EndWhile
    
    \State Update $B_{i+1} \gets B_i - t_i \nabla f(B_i) $.
    \State Update $i \gets i + 1$.
\EndWhile
\State \Return $B^{*} = B_i$.
\end{algorithmic}
\end{algorithm}

Notice that in Problem \eqref{sub1}, the LQG performance $J_{\text{LQG}}$ is nonconvex but smooth, while the configuration cost $\Phi(B)$ is assumed to be convex but not necessarily smooth. Therefore, direct optimization of the overall objective function would lead to a nonconvex and nonsmooth optimization problem, which poses significant challenges for algorithm design. To address these difficulties, we utilize the celebrated alternating direction method of multipliers (ADMM), which allows us to handle the nonconvex smooth term and the convex nonsmooth term in the objective function separately. By introducing an auxiliary variable $M$, we can reformulate Problem \eqref{sub1} as
\begin{equation}\label{accept}
    \begin{split}
       & \min_{B\in\mathcal{S}, M}~  J_{\text{LQG}}(B, C_h) + \gamma\Phi(M) \\
&\text{subject to}~  B = M,~M \in \Omega_B.
    \end{split}
\end{equation}
The augmented Lagrangian dual function for this constrained optimization problem is given by
\begin{equation}
\begin{split}
    \mathcal{L}_\rho(B, M, L_B) = &J_{\text{LQG}}(B, C_h) + \gamma\Phi(M) +I_{\Omega_B}(M) \\
    &~~~+ \langle L_B, B-M \rangle + \frac{\rho}{2}\|B-M\|_{\F}^2,
\end{split}
\end{equation}
where $L_B \in \mathbb{R}^{n \times m}$ represents the Lagrange multiplier matrix and $\rho > 0$ denotes the penalty parameter. Following \cite{boyd2004convex}, the standard ADMM consists of the following iterations:
\begin{equation*}
\begin{split}
    &B^{k+1} = \argmin_{B \in \mathcal{S}} J_{\text{LQG}}(B, C_h) + \langle L_B^k, B \rangle + \frac{\rho}{2}\|B-M^k\|_{\F}^2,\\
    &M^{k+1} = \argmin_{M\in \Omega_B} \gamma\Phi(M) - \langle L_B^k, M \rangle + \frac{\rho}{2}\|B^{k+1}-M\|_{\F}^2,\\
    &L_B^{k+1} = L_B^k + \rho(B^{k+1} - M^{k+1}),
\end{split}
\end{equation*}
and terminates until both the primal and dual residuals satisfy the following numerical conditions:
\begin{equation*}
    \|B^{k+1} - M^{k+1}\|_{\F} \leq \epsilon^{\text{pri}},~\rho\|M^{k+1} - M^k\|_{\F} \leq \epsilon^{\text{{dual}}},
\end{equation*}
where $\epsilon^{\text{pri}}$ and $\epsilon^{\text{{dual}}}$ denote the predetermined primal and dual feasibility tolerances. 
Notice that updating the auxiliary variable $M$ involves solving a convex optimization problem, which often exhibits explicit-form solutions when the configuration cost $\Phi(\cdot)$ and constraint set $\Omega_B$ have special structures. Although the LQG performance $J_{\text{LQG}}(B,C_h)$ is nonconvex, updating $B$ only involves solving a smooth optimization problem, and the additional quadratic term helps convexify the objective function when $\rho$ is sufficiently large. In practical implementation, we can update $B$ using the gradient descent method with backtracking line search, as shown in Algorithm 1, where at least a stationary point can be obtained \cite{nocedal1999numerical}. For simplicity, we denote $f(B):=\mathcal{L}_\rho(B, M^k, L_B^k)$ in Algorithm 1. Note that the gradient descent is not a unique method, and here we only present an example with backtracking line search. Please refer to \cite{nocedal1999numerical} for more decent methods.

Now we have analyzed the subproblem in \eqref{sub1} and developed a computational method based on ADMM. The analysis and computation of \eqref{sub2} are similar and omitted. The alternating optimization scheme for the sensor and actuator configuration problem is then summarized in Algorithm 2.

\begin{algorithm}[!]
\caption{Optimization-based Sensor and Actuator Configuration in LQG Control Systems}
\label{alg:alternating}
\begin{algorithmic}[1]
    \State Initialize $B_0 \in \Omega_B$ and $C_0 \in \Omega_C$.
    \State Set tolerances $\epsilon^{\text{pri}},\epsilon^{\text{{dual}}}$ and $\epsilon^{\text{{main}}}$.
    \State Set parameter $\rho > 0$ and the iterator $h \gets 0$.
    \While{not convergent}
        \State Set the iterator $k \gets 0$ and fix $C = C_h$.
        \State Initialize auxiliary and dual variables: $M^0,L_{B}^0$.
        \While{not convergent}
            \State Update $B^{k+1}$ using gradient descent method.
            \State Update $M^{k+1}$ by solving 
            \begin{align*}
                ~~~~~~~~~\min_{M\in\Omega_B} \gamma\Phi(M) - \langle L_B^k, M \rangle + \frac{\rho}{2}\|B^{k+1}-M\|_{\F}^2.
            \end{align*}
            \State Update $L_B^{k+1} = L_B^k + \rho(B^{k+1} - M^{k+1})$.
            \State Check ADMM convergence conditions
            \begin{equation*}
~~~~~~~~~\begin{cases}
    \|B^{k+1} - M^{k+1}\|_{\F} \leq \epsilon^{\text{pri}}, \\\rho\|M^{k+1} - M^k\|_{\F} \leq \epsilon^{\text{{dual}}}.
\end{cases}
\end{equation*}
 \State Update $k \gets k+1$.
        \EndWhile
        
        \State Update the actuator matrix $B_{h+1} = M^{k}$.
       \State Set the iterator $k \gets 0$ and fix $B = B_{h+1}$.
        \State Initialize auxiliary and dual variables: $N^0,L_{C}^0$.
        \While{not convergent}
            \State Update $C^{k+1}$ using gradient descent method.
            \State Update $N^{k+1}$ by solving 
            \begin{align*}
                ~~~~~~~~~\min_{N\in\Omega_C} \gamma\Psi(N) - \langle L_C^{k}, N \rangle + \frac{\rho}{2}\|C^{k+1}-N\|_{\F}^2.
            \end{align*}
            \State Update $L_C^{k+1} = L_C^{k} + \rho(C^{k+1} - N^{k+1})$.
            \State Check ADMM convergence conditions
            \begin{equation*}
~~~~~~~~~\begin{cases}
    \|C^{k+1} - N^{k+1}\|_{\F} \leq \epsilon^{\text{pri}},\\
    \rho\|N^{k+1} - N^{k}\|_{\F} \leq \epsilon^{\text{{dual}}}.
\end{cases}
\end{equation*}
 \State Update $k \gets k+1$.
        \EndWhile
        \State Update the sensor matrix $C_{h+1} = N^{k}$.
        \State Update $h \gets h + 1$.
        \State Check outer loop convergence
        \begin{equation*}
    ~~~~~~~~~\|B_{h+1} - B_{h}\|_{\F} +\|C_{h+1} - C_{h}\|_{\F}  \leq \epsilon^{\text{{main}}}.
\end{equation*}
    \EndWhile
    \State \Return $(B^*, C^*) = (B_h, C_h)$.
\end{algorithmic}
\end{algorithm}

\begin{myrem}
When the penalty parameter $\rho > 0$ is chosen sufficiently large, the augmented quadratic term in the ADMM is dominant such that the overall optimization is convexified, and ADMM converges to a stationary point \cite{wang2019global}. Furthermore, it is worth noting that the convergence of Algorithm 2 follows from the monotonically nonincreasing property of the bi-level alternating optimization scheme, i.e., $\forall h\ge 1$,
\begin{align}
J(B_{h+1},C_{h}) \le J(B_{h}, C_{h}) \le J(B_{h}, C_{h-1}).
\end{align}
Although the global optimality cannot be guaranteed due to the nonconvexity of $J_{\text{LQG}}$, we will show that at least a stationary solution can be obtained using Algorithm 2.
\end{myrem}
\vspace{3pt}

% \begin{myrem}
%     It is worth noting that we use auxiliary variables ($M, N$) rather than original variables ($B, C$) in the main-loop updates, i.e., $B_{h+1} = M^{k+1}$ and $C_{h+1} = N^{k+1}$. While one might intuitively expect the auxiliary variables ($M, N$) to converge to the same values as the original variables ($B, C$) within the ADMM framework, their numerical characteristics differ in implementations. The justification for this distinction and its implications are revealed in the following theorem.
% \end{myrem}
% \vspace{3pt}
\begin{mythr}
    Let $B^* = \lim_{h\rightarrow\infty} B_h$ and $C^* = \lim_{h\rightarrow\infty} C_h$ denote the accumulation points of the alternating optimization scheme in Algorithm 2. Given that the numerical tolerances $\epsilon^{\text{pri}},\epsilon^{\text{{dual}}}$ and $\epsilon^{\text{{main}}}$ are sufficiently small, $(B^*,C^*)$ is at least a stationary point to the optimization problem in \eqref{p0}.
\end{mythr}
\vspace{3pt}
Proof. By the numerical property of the backtracking line search method used in Algorithm 1, we first have that $B^{k+1}$ is a stationary point to the following optimization problem:
\begin{align}\label{th41}
    \min_{B} J_{\text{LQG}}(B, C^*) + \langle L_B^k, B \rangle + \frac{\rho}{2}\|B-M^k\|_{\F}^2.
\end{align}
Based on the first-order optimality condition of \eqref{th41}, we can derive the following condition:
\begin{equation}\label{th42}
           0\in \frac{\partial J_{\text{LQG}}(B,C^*)}{\partial B}\Bigr\rvert_{B=B^{k+1}}+L_B^k+\rho (B^{k+1} - M^k).
\end{equation}
Meanwhile, by the first-order optimality condition of the $M$-update in the ADMM, we have that 
\begin{equation}\label{th43}
   \begin{split}
       0&\in \gamma\frac{\partial \Phi(M)}{\partial M}\Bigr\rvert_{M=M^{k+1}} -L_B^k \\
       &~~~~~~~~~~~~~~~-\rho (B^{k+1} - M^{k+1})+ \mathcal{N}_{\Omega_B}(M^{k+1}).
   \end{split}
\end{equation}
Combining \eqref{th42} and \eqref{th43}, we further obtain that
\begin{equation}\label{th45}
    \begin{split}
           0&\in \frac{\partial J_{\text{LQG}}(B,C^*)}{\partial B}\Bigr\rvert_{B=B^{k+1}}+\gamma\frac{\partial \Phi(M)}{\partial M}\Bigr\rvert_{M=M^{k+1}} \\
   &~~~~~~~~~~~~~~~~~~~+\rho(M^{k+1} - M^k)+\mathcal{N}_{\Omega_B}(M^{k+1}).
    \end{split}
\end{equation}
By Theorem 1, we have that
\begin{align}
    \frac{\partial J_{\text{LQG}}}{\partial B} = -2 P(G_1+G_2)PBR^{-1}
\end{align}
is a continuous function with respect to $B$, which follows from the continuity of Lyapunov equations. Since $\|B^{k+1} - M^{k+1}\|_{\F} \leq \epsilon^{\text{pri}}$, we can obtain that 
\begin{align}\label{th46}
    \frac{\partial J_{\text{LQG}}(B,C^*)}{\partial B}\Bigr\rvert_{B=B^{k+1}} = \frac{\partial J_{\text{LQG}}(B,C^*)}{\partial B}\Bigr\rvert_{B=M^{k+1}}
\end{align}
as $\epsilon^{\text{pri}} \rightarrow 0$. Meanwhile, we have that
\begin{align}\label{th47}
    \rho(M^{k+1} - M^k) = 0\text{ as }\epsilon^{\text{{dual}}} \rightarrow 0,
\end{align}
which follows from $\rho\|M^{k+1} - M^k\|_{\F} \leq \epsilon^{\text{{dual}}}$. Combing \eqref{th46} and \eqref{th47}, and notice that $B_{h+1} = M^{k+1}$, we obtain that
\begin{align}
    0 \in \gamma\frac{\partial \Phi(B)}{\partial B} + \mathcal{N}_{\Omega_B}(B) - 2 P(G_1+G_2)PBR^{-1}
\end{align}
holds at point $B = B^*$. Similarly, we can also obtain that
\begin{align}
    0\in \gamma\frac{\partial \Psi(C)}{\partial C} + \mathcal{N}_{\Omega_C}(C)-2\Pi_v^{-1}CX(H_1+H_2)X
\end{align}
holds at point $C = C^*$.
By Theorem 3, we finally conclude that $(B^*,C^*)$ satisfies the first-order optimality condition and is at least a stationary point to Problem \eqref{p0}.
$\hfill \square$

\subsection{Scenario I: Sparsity Promoting Configuration}
In this section, we consider the sparsity promoting configuration of sensors and actuators, where 
\begin{equation}
    \begin{split}
          &\Phi(B) = \|B\|_0,~\Psi(C) = \|C\|_0,\\
    &~~\Omega_B = \mathbb{R}^{n\times m},~\Omega_C=\mathbb{R}^{q\times n}.
    \end{split}
\end{equation}
This setup can be regarded as a specific application scenario of the problem formulated in Section II. Let us consider a large-scale networked control system in \eqref{sys}, where each element $[x(t)]_i$ represents the state of a subsystem.
Notice that each nonzero element $[B]_{ij}$ represents an actuating channel from actuator $j$ to subsystem $i$, and each nonzero element $[C]_{ij}$ represents a sensing channel from subsystem $j$ to sensor $i$. To reduce the total number of actuating and sensing channels, we can define the configuration cost as follows:
\begin{align}
   J_{\text{SAC}} = \|B\|_0+\|C\|_0.
\end{align}
Due to the nonconvexity of the $\ell_0$ norm, it is typical to use the $\ell_1$ norm to promote the sparsity in many applications, which is well-known as the $\ell_1$ relaxation \cite{donoho2003optimally}. Another advantage of using the $\ell_1$ norm is that it can help suppress the magnitudes of the elements of sensor and actuator matrices. Therefore we instead consider the following configuration cost:
\begin{align}
   \tilde{J}_{\text{SAC}} = \|B\|_1+\|C\|_1,
\end{align}
which is a convex but nonsmooth function regarding matrices $B$ and $C$. More specifically, the sparsity promoting configuration (SPC) problem can be formulated as follows:
\begin{align}\label{s1}\tag{S1}
    \min_{B,C}J_{\text{LQG}}(B,C)+ \gamma(\|B\|_1+\|C\|_1). 
\end{align}
Notice that the above formulation can also be easily extended to include block sparsity, while here we assume the dimension of each subsystem to be $1$ for simplicity.

Based on the optimization framework established in Section III.A, we derive customized versions of the first-order optimality condition and the alternating optimization algorithm for the sparsity promoting configuration problem in \eqref{s1}.
\vspace{3pt}

\begin{mycor}
 If a matrix pair $(B,C)$ is an optimal solution to Problem \eqref{s1}, it must satisfy the following conditions:
\begin{equation}\label{c11m}
     \begin{cases}
        [P(G_1+G_2)PBR^{-1}]_{ij} = -\frac{\gamma}{2}, & \text{if }[B]_{ij} <0,\\
    -\frac{\gamma}{2}\le [P(G_1+G_2)PBR^{-1}]_{ij} \le \frac{\gamma}{2}, & \text{if }[B]_{ij} =0,\\
    [P(G_1+G_2)PBR^{-1}]_{ij} = \frac{\gamma}{2}, & \text{if }[B]_{ij} >0,\\
   \end{cases}
\end{equation}
    and
\begin{equation}\label{c12m}
     \begin{cases}
        [\Pi_v^{-1}CX(H_1+H_2)X]_{ij} = -\frac{\gamma}{2}, & \text{if }[C]_{ij} <0,\\
    -\frac{\gamma}{2}\le [\Pi_v^{-1}CX(H_1+H_2)X]_{ij} \le \frac{\gamma}{2}, & \text{if }[C]_{ij} =0,\\
    [\Pi_v^{-1}CX(H_1+H_2)X]_{ij} = \frac{\gamma}{2}, & \text{if }[C]_{ij} >0.\\
   \end{cases}
\end{equation}
Conversely, if a matrix pair $(B,C)$ satisfies the above conditions, it is at least a stationary point to Problem \eqref{s1}.
\end{mycor}
\vspace{3pt}
Proof. Given that $\Omega_B = \mathbb{R}^{n \times m}$ and $\Omega_C = \mathbb{R}^{q \times n}$, by the definition of normal cones, we have that $\mathcal{N}_{\Omega_B}(B) = \emptyset$ and $\mathcal{N}_{\Omega_C}(C) = \emptyset$. Therefore, the first-order optimality conditions in Theorem 3 can be reduced to
\begin{equation}\label{c11}
    \begin{split}
                &2 P(G_1+G_2)PBR^{-1} \in \gamma\frac{\partial \Phi(B)}{\partial B},\\
        &2\Pi_v^{-1}CX(H_1+H_2)X\in \gamma\frac{\partial \Psi(C)}{\partial C}.
    \end{split}
\end{equation}
Notice that the $\ell_1$ norm is a nonsmooth function, which is non-differentiable at the point $0$. However, it admits subdifferential. Specifically, for any scalar $x \in \mathbb{R}$, we have
\begin{equation}\label{c12}
    \begin{cases}
        {\partial |x|}/{\partial x} = 1, &\text{if } x>0,\\
    {\partial |x|}/{\partial x} \in [-1,1], &\text{if } x=0,\\
    {\partial |x|}/{\partial x} = -1, &\text{if } x<0.\\
\end{cases}
\end{equation}
Since $\Phi(B) = \|B\|_1 = \sum_{i,j}|[B]_{ij}|$ and $\Psi(C)=\|C\|_1=\sum_{i,j}|[C]_{ij}|$, the optimality conditions in \eqref{c11m} and \eqref{c12m} can be derived elementwisely by combining \eqref{c11} and \eqref{c12}. $\hfill \square$
\vspace{3pt}

When using Algorithm 2 to solve Problem \eqref{s1}, most steps remain standard and irrelevant to the choices of configuration cost functions, except the $M$-update and $N$-update steps:
\begin{align}
    &\min_{M\in\Omega_B} \gamma\Phi(M) - \langle L_B^k, M \rangle + \frac{\rho}{2}\|B^{k+1}-M\|_{\F}^2,\label{mup}\tag{$M$-update}\\
    &\min_{N\in\Omega_C} \gamma\Psi(N) - \langle L_C^{k}, N \rangle + \frac{\rho}{2}\|C^{k+1}-N\|_{\F}^2.\label{nup}\tag{$N$-update}
\end{align}
The explicit-form solutions for the $M$-update and $N$-update steps can be obtained for the sparsity promoting configuration problem through the following lemma.
\vspace{3pt}

\begin{mylem}
    Regarding Problem \eqref{s1}, the optimal solutions for the $M$-update and $N$-update steps are given by
   \begin{equation}
       \begin{cases}
            M^{k+1} = \text{Soft}_{\gamma/\rho}(B^{k+1} + \frac{1}{\rho}L_B^k),\\
        N^{k+1} = \text{Soft}_{\gamma/\rho}(C^{k+1} + \frac{1}{\rho}L_C^{k}),
       \end{cases}
   \end{equation}
    where $\text{Soft}_{\tau}(\cdot)$ is known as the soft thresholding operator and is defined such that for any matrix $X$,
    \begin{align}
        [\text{Soft}_{\tau}(X)]_{ij} = \begin{cases}
            [X]_{ij} - \tau, & \text{if }[X]_{ij} \ge \tau,\\
            0, & \text{if }-\tau<[X]_{ij}<\tau,\\
            \tau+[X]_{ij}, & \text{if }[X]_{ij} \le -\tau.\\
        \end{cases}
    \end{align}
\end{mylem}
\vspace{5pt}
Proof. Notice that $\Omega_B = \mathbb{R}^{n\times m}$ and the $M$-update step is equivalent to solving the following problem:
\begin{align}
    \min_{M\in\mathbb{R}^{n\times m}} \Phi(M)  + \frac{\rho}{2\gamma}\|B^{k+1}-M+\frac{1}{\rho}L_B^k\|_{\F}^2.
\end{align}
Since $\Phi(M) = \|M\|_1 = \sum_{i,j} |[M]_{ij}|$, the above optimization problem can be solved elementwisely by
\begin{align}\label{th51}
    \min_{[M]_{ij}} |[M]_{ij}|  + \frac{\rho}{2\gamma}([B^{k+1}+\frac{1}{\rho}L_B^k]_{ij} - [M]_{ij})^2
\end{align}
for any indices $1\le i\le n,1\le j\le m $. By the results in (\S 23, \cite{rockafellar1997convex}), the solution to \eqref{th51} can be obtained using the soft thresholding operator, i.e., 
\begin{align}
    [M]_{ij} = \text{Soft}_{\gamma/\rho}([B^{k+1} + \frac{1}{\rho}L_B^k]_{ij}).
\end{align}
The optimal solution for the $N$-update step can be derived in a similar way and is omitted here. $\hfill \square$

\subsection{Scenario II: Low-Rank Promoting Configuration}
In this section, we consider the low-rank promoting configuration of sensors and actuators, where 
\begin{equation}
    \begin{split}
         \Phi(B) &= \rank(B),~\Psi(C) = \rank(C),\\
    &\Omega_B = \mathbb{R}^{n\times m},~\Omega_C=\mathbb{R}^{q\times n}.
    \end{split}
\end{equation}
This setup can be regarded as a specific application scenario of the problem formulated in Section II. We first introduce the following well-known rank factorization lemma \cite{lay2003linear}.
\vspace{3pt}

\begin{mylem}
    For any matrix $A\in\mathbb{R}^{m \times n}$ whose rank is $r$, there exist $X \in \mathbb{R}^{m \times r}$ and $Y \in \mathbb{R}^{r\times n}$ such that $A = XY$.
\end{mylem}
\vspace{3pt}

Notice that if the actuator matrix $B\in \mathbb{R}^{n\times m}$ has a low rank $r_B<m$, it admits the following factorization:
\begin{align}
B = B_1B_2,~B_1\in \mathbb{R}^{n\times r_B},B_2 \in \mathbb{R}^{r_B\times m},
\end{align}
which implies that $ Bu(t) = B_1(B_2u(t))$.
Consequently, the actuator matrix can be reduced to $\tilde{B} = B_1$ with $r_B$ columns, with an intermediate signal $\tilde{u}(t)=B_2u(t)\in \mathbb{R}^{r_B}$. This reduction allows the number of actuators to decrease from $m$ to $r_B$. Similarly, if the sensor matrix $C\in \mathbb{R}^{q\times n}$ has a low rank $r_C < q$, we can do the following factorization:
\begin{align}
C = C_1C_2,~C_1\in \mathbb{R}^{q\times r_C},~C_2 \in \mathbb{R}^{r_C\times n},
\end{align}
which implies that $Cx(t) = C_1(C_2x(t))$. The sensor matrix can be reduced to $\tilde{C} = C_2$ with $r_C$ rows, with an intermediate measurement $\tilde{y}(t)=C_2x(t)\in \mathbb{R}^{r_C}$. Through this transformation, the number of sensors can be reduced from $q$ to $r_C$.

To reduce the total number of used actuators and sensors, we can define the configuration cost as follows:
\begin{align}
   J_{\text{SAC}} = \rank(B)+\rank(C).
\end{align}
Due to the nonconvexity and discontinuity of matrix ranks, the nuclear norm is commonly applied to promote the low-rank property in numerical optimization \cite{donoho2003optimally}. For any matrix $A\in \mathbb{R}^{m\times n}$, its nuclear norm is defined as the sum of singular values, i.e., $\|A\|_{*}= \sum_{i} \sigma_i(A)$.
Therefore, we instead consider the following configuration cost:
\begin{align}
   \tilde{J}_{\text{SAC}} = \|B\|_{*}+\|C\|_{*},
\end{align}
which is a convex but nonsmooth function regarding matrices $B$ and $C$. More specifically, the low-rank promoting configuration (LPC) problem can be formulated as follows:
\begin{align}\label{s2}\tag{S2}
    \min_{B,C}J_{\text{LQG}}(B,C)+ \gamma(\|B\|_{*}+\|C\|_{*}). 
\end{align}
Based on the optimization framework established in Section III.A, we derive customized versions of the first-order optimality condition and the alternating optimization algorithm for the low-rank promoting configuration problem in \eqref{s2}.
\vspace{3pt}

\begin{mylem}
    The subdifferential of the nuclear norm $\|\cdot\|_{*}$ at a matrix $A$, with SVD $A = U\Sigma V^{\T}$, is given by
    \begin{align*}
        \frac{\partial \|A\|_{*}}{\partial A} = \{UV^{\T}+W\mid U^{\T}W=0,WV = 0,\|W\|_{2}\le 1\}.
    \end{align*}
\end{mylem}
\vspace{7pt}
Proof. It follows from the results in \cite{watson1992characterization}. $\hfill \square$
\vspace{3pt}

\begin{mycor}
     If a matrix pair $(B,C)$ is an optimal solution to Problem \eqref{s2}, it must satisfy the following conditions:
     \begin{equation}\label{c21m}
         \begin{cases}
               \| 2 P(G_1+G_2)PBR^{-1} -\gamma U_BV_B^{\T} \|_{2}\le \gamma,\\
        2 U_B^{\T}P(G_1+G_2)PBR^{-1} =\gamma V_B^{\T},\\
        2 P(G_1+G_2)PBR^{-1}V_B = \gamma U_B,
         \end{cases}
     \end{equation}
     and
     \begin{equation}\label{c22m}
         \begin{cases}
               \| 2\Pi_v^{-1}CX(H_1+H_2)X - \gamma U_CV_C^{\T} \|_{2}\le \gamma,\\
        2 U_C^{\T}\Pi_v^{-1}CX(H_1+H_2)X = \gamma V_C^{\T},\\
        2 \Pi_v^{-1}CX(H_1+H_2)XV_C =\gamma U_C,
         \end{cases}
     \end{equation}
     where $U_B,U_C,V_B,V_C$ represent the left and right singular vector matrices for $B$ and $C$, respectively, i.e.,
    \begin{align}
        B = U_B\Sigma_BV_B^{\T}\text{~~~and~~~}C = U_C\Sigma_CV_C^{\T}.
    \end{align}
    Conversely, if a matrix pair $(B,C)$ satisfies the above conditions, it is at least a stationary point to Problem \eqref{s2}.
\end{mycor}
\vspace{7pt}
Proof. By Theorem 3, the first-order optimality conditions for Problem \eqref{s2} can be expressed as
\begin{align}
                    &2 P(G_1+G_2)PBR^{-1} \in\gamma \frac{\partial \Phi(B)}{\partial B}, \label{c21} \\
        &2\Pi_v^{-1}CX(H_1+H_2)X\in \gamma\frac{\partial \Psi(C)}{\partial C}.\label{c2c}
\end{align}
By expanding the subdifferential ${\partial \Phi(B)}/{\partial B}$ using Lemma 3, we can further conclude that \eqref{c21} holds if and only if there exists some $W_B$ such that 
\begin{align}
    &U_B^{\T}W_B = 0,~W_BV_B=0,~\|W_B\|_{2}\le 1,\label{c22}\\
    &2 P(G_1+G_2)PBR^{-1} = \gamma(U_BV_B^{\T}+W_B).\label{c23}
\end{align}
From \eqref{c23}, by taking $W_B = 2 P(G_1+G_2)PBR^{-1}-U_BV_B^{\T}$ and substituting it into \eqref{c22}, we obtain the conditions in \eqref{c21m}. Based on the optimality condition in \eqref{c2c}, the proof of \eqref{c22m} can be derived similarly and is omitted here.$\hfill \square$
\vspace{3pt}

The above corollary provides a computational method for evaluating the optimality of numerical solutions using singular value decomposition. In the following lemma, we develop the explicit-form solutions for the $M$-update and $N$-update steps when using Algorithm 2 to solve Problem \eqref{s2}.
\vspace{3pt}

\begin{mylem}
    Regarding Problem \eqref{s2}, the optimal solutions for the $M$-update and $N$-update steps are given by
   \begin{equation}
       \begin{cases}
            M^{k+1} = \text{SVT}_{\gamma/\rho}(B^{k+1} + \frac{1}{\rho}L_B^k),\\
        N^{k+1} = \text{SVT}_{\gamma/\rho}(C^{k+1} + \frac{1}{\rho}L_C^{k}),
       \end{cases}
   \end{equation}
    where $\text{SVT}_{\tau}(\cdot)$ is known as the singular value thresholding (SVT) operator. For any matrix $X$ with singular value decomposition $X = U\Sigma V^{\T}$, the SVT operator is defined as
    \begin{align}\label{lem41m}
        \text{SVT}_{\tau}(X) = U(\Sigma-\tau I)_+V^{\T}.
    \end{align}
\end{mylem}
\vspace{5pt}
Proof.
Since the $M$-update and $N$-update steps have the same form, we only need to prove the case for $M$-update:
\begin{equation}
    \min_{M\in\Omega_B}\gamma \Phi(M) - \langle L_B^k, M \rangle + \frac{\rho}{2}\|B^{k+1}-M\|_{\F}^2,
\end{equation}
where $\Phi(M) = \|M\|_*$ is the nuclear norm. Let $X = B^{k+1} + \frac{1}{\rho}L_B^k$, then the $M$-update step can be rewritten as
\begin{equation}\label{lem41}
  M^{k+1}=  \min_{M\in\mathbb{R}^{n\times m}} \gamma\|M\|_* + \frac{\rho}{2}\|M-X\|_{\F}^2.
\end{equation}
Since \eqref{lem41} is a convex optimization problem, $M^{k+1}$ is an optimal solution if and only if 
\begin{equation}\label{lem42}
    0 \in \partial\|M^{k+1}\|_* + \frac{\rho}{\gamma}(M^{k+1}-X).
\end{equation}
Hence, it suffices to show that $M^{k+1} =  \text{SVT}_{\gamma/\rho}(X)$ satisfies the above condition. Let $U_{+}$ and $V_{+}$ denote the sub-matrices of $U,V$ corresponding to singular values $\sigma_i(X)\ge\frac{\gamma}{\rho}$, then the singular value decomposition of $M^{k+1}$ can be obtained as
\begin{align*}
    M^{k+1} = U_M\Sigma_MV_M^{\T} \text{ with } U_M = \begin{bmatrix}
        U_+ & 0
    \end{bmatrix}, V_M = \begin{bmatrix}
        V_+ \\ 0
    \end{bmatrix},
\end{align*}
and $\Sigma_M = (\Sigma-\frac{\gamma}{\rho} I)_+$.
By Lemma 3, \eqref{lem42} can be rewritten as
\begin{align}\label{lem43}
    U_MV_M^{\T} + W + \frac{\rho}{\gamma}(M^{k+1}-X) = 0
\end{align}
for some satisfactory $W$. By Eqn \eqref{lem41m}, we have that
\begin{align}
   M^{k+1}-X &= U_M\Sigma_MV_M^{\T} - U\Sigma V^{\T}\\
   &= -\frac{\gamma}{\rho} U_MV_M^{\T}-\begin{bmatrix}
         0 & U_-
    \end{bmatrix}\Sigma\begin{bmatrix}
        0\\V_-
    \end{bmatrix},
\end{align}
where $U_{-}$ and $V_{-}$ denote the sub-matrices of $U,V$ corresponding to singular values $\sigma_i(X)<\frac{\gamma}{\rho}$. By \eqref{lem43}, we obtain that
\begin{align*}
    W = -\frac{\rho}{\gamma}( M^{k+1}-X)-U_MV_M^{\T} = \frac{\rho}{\gamma} \begin{bmatrix}
         0 & U_-
    \end{bmatrix}\Sigma\begin{bmatrix}
        0\\V_-
    \end{bmatrix}.
\end{align*}
As $U,V$ are unitary matrices, it is easy to verify that $U_M^{\T}W = 0, WV_M = 0$ and $\|W\|_{2}\le 1$. Therefore \eqref{lem42} holds. $\hfill\square$

\subsection{Scenario III: Structure-Constrained Configuration}
In this section, we consider the structure-constrained configuration of sensors and actuators, where 
\begin{equation}\label{scp1}
    \begin{split}
    &\Omega_B = \{B\in\mathbb{R}^{n\times m}\mid B \circ S_B = 0\},\\
    &\Omega_C=\{C\in\mathbb{R}^{q\times n}\mid C\circ S_C = 0\},\\
      &\Phi(B) = \|B\|_{\text{F}}^2,~\Psi(C) = \|C\|_{\text{F}}^2,
    \end{split}
\end{equation}
where $S_B \in \mathbb{R}^{n\times m}$ and $S_C\in \mathbb{R}^{q\times n}$ are prescribed structure matrices such that $[S_B]_{ij}\in \{0,1\}$ and $[S_C]_{ij}\in \{0,1\}$. Let us consider a large-scale networked control system in \eqref{sys}, where each element $[x(t)]_i$ represents the state of a subsystem.
Notice that if $[S_B]_{ij}=1$, it means that the $i$-th subsystem cannot be actuated by the $j$-th actuator. Similarly, if $[S_C]_{ij}=1$, it means that the $j$-th subsystem cannot be measured by the $i$-th sensor.
This setup describes the hardware constraints of networked control systems and can be regarded as a specific application scenario of the problem formulated in Section II. More specifically, the structure-constrained configuration (SCC) problem can be formulated as follows:
\begin{equation}\label{s3}\tag{S3}
    \begin{split}
        \min_{B,C}~J_{\text{LQG}}&(B,C)+\gamma(\|B\|_{\text{F}}^2+\|C\|_{\text{F}}^2) \\
        \text{ subject to}&~B \circ S_B = 0,\\
        &~C\circ S_C = 0.
    \end{split}
\end{equation}
Based on the optimization framework established in Section III.A, we derive customized versions of the first-order optimality condition and the alternating optimization algorithm for the structure-constrained configuration problem in \eqref{s3}.
\vspace{3pt}

\begin{mycor}
 If a matrix pair $(B,C)$ is an optimal solution to Problem \eqref{s3}, it must satisfy the following conditions:
\begin{align}
            &[P(G_1+G_2)PBR^{-1} -\gamma B]\circ \bar{S}_B = 0,\label{c31m}\\
   &[\Pi_v^{-1}CX(H_1+H_2)X-\gamma C] \circ \bar{S}_C = 0.\label{c32m}
\end{align}
where $\bar{S}$ denotes the complement of a binary matrix $S$ such that $[\bar{S}]_{ij}+[S]_{ij}=1$.
Conversely, if $(B,C)$ satisfies the above conditions, it is at least a stationary point to Problem \eqref{s3}.
\end{mycor}
\vspace{3pt}
Proof.  By Theorem 3, the first-order optimality conditions for Problem \eqref{s3} can be expressed as
\begin{align}
                    &2 P(G_1+G_2)PBR^{-1} -2\gamma B\in \mathcal{N}_{\Omega_B}, \label{c31} \\
        &2\Pi_v^{-1}CX(H_1+H_2)X-2\gamma C\in \mathcal{N}_{\Omega_C}.\label{c32}
\end{align}
By the definition of normal cones, we have that $\Lambda_B \in \mathcal{N}_{\Omega_B}$ if and only if for any $B' \in \Omega_{B}$, i.e., $B'\circ S_B=0$,
\begin{align}
   \tr(\Lambda_B^{\T} (B' - B)) \le 0.
\end{align}
As $(B,C)$ is a feasible point and $B\in \Omega_{B}$, we can conclude that $\Lambda_B \circ \bar{S}_B = 0$ and therefore
\begin{align}
    \mathcal{N}_{\Omega_B} = \{\Lambda_B\in \mathbb{R}^{n\times m}\mid \Lambda_B\circ \bar{S}_B = 0\}.
\end{align}
The condition in \eqref{c31} can be rewritten as \eqref{c31m}. The proof of \eqref{c32m} can be obtained similarly and is omitted here. $\hfill\square$
\vspace{3pt}

The above corollary shows that the optimal solution and its first-order gradient have complementary structures. In the following lemma, we develop the explicit-form solutions for the $M$-update and $N$-update steps for Problem \eqref{s3}.
\vspace{3pt}

\begin{mylem}
     Regarding Problem \eqref{s3}, the optimal solutions for the $M$-update and $N$-update steps are given by
      \begin{equation}\label{lem51m}
       \begin{cases}
            M^{k+1} = \frac{1}{\rho+2\gamma}(L_B^k+\rho B^{k+1})\circ \bar{S}_B,\\
        N^{k+1} = \frac{1}{\rho+2\gamma}(L_C^k+\rho C^{k+1})\circ \bar{S}_C.
       \end{cases}
   \end{equation}
\end{mylem}
\vspace{5pt}
Proof. Given that $\Phi(B) = 0$ and $\Omega_B = \{B\in\mathbb{R}^{n\times m}\mid B \circ S_B = 0\}$, the $M$-update step for Problem \eqref{s3} is solving the following constrained optimization problem:
\begin{equation}\label{lem51}
    \min_{M\circ S_B = 0} \gamma\|M\|_{\text{F}}^2 - \langle L_B^k, M \rangle + \frac{\rho}{2}\|B^{k+1}-M\|_{\F}^2.
\end{equation}
It is easy to see that if $M\circ S_B = 0$, 
\begin{align*}
  \|B^{k+1}-M\|_{\F}^2 = \|B^{k+1}\circ \bar{S}_B-M\|_{\F}^2+\|B^{k+1}\circ {S}_B\|_{\F}^2.
\end{align*}
Furthermore, the following equality holds:
\begin{align*}
    \argmin_{M\circ S_B = 0} \|B^{k+1}-M\|_{\F}^2 = \argmin_{M} \|B^{k+1}\circ \bar{S}_B-M\|_{\F}^2.
\end{align*}
This is because the optimal solution to the right-hand side optimization will naturally satisfy the constraint $M\circ S_B = 0$. Meanwhile, notice that if $M\circ S_B = 0$,
\begin{align}
    \langle L_B^k, M \rangle = \langle L_B^k\circ \bar{S}_B, M \rangle = \langle L_B^k, M\circ \bar{S}_B \rangle.
\end{align}
Therefore, we can conclude that \eqref{lem51} is equivalent to solving the following unconstrained optimization problem:
\begin{align*}
    \min_{M} \gamma \|M\|_{\text{F}}^2- \langle L_B^k\circ \bar{S}_B, M \rangle + \frac{\rho}{2}\|B^{k+1}\circ \bar{S}_B-M\|_{\F}^2.
\end{align*}
Its solution can be obtained using the completion of squares, as shown in \eqref{lem51m}. The solution for the $N$-update step can be obtained similarly and the derivation is omitted here. $\hfill \square$

\section{Simulations}
In this section, we present numerical simulations to demonstrate the effectiveness of our proposed methods. The experiments were performed using MATLAB R2023a on a MacBook Air (2024) equipped with 16 GB memory and an M3 CPU.

We consider the benchmark example REA1 from the depository COMPleib \cite{leibfritz2003description}, which describes the dynamics of a chemical reactor. Since we are solving the sensor and actuator configuration problem, we assume only the system matrix $A$ of system \eqref{sys} is known and fixed as follows:
\begin{align*}
     A = \begin{bmatrix}
    1.38& -0.2077& 6.715& -5.676\\
     -0.5814& -4.29& 0& 0.675\\
     1.067& 4.273& {-6.654}& 5.893\\
     0.048& 4.273& 1.343& -2.104
    \end{bmatrix},
\end{align*}
while the actuator matrix $B\in\mathbb{R}^{4\times 4}$ and the sensor matrix $C\in \mathbb{R}^{4\times 4}$ are unknown and to be determined\footnote{For illustration purposes, here we assume the dimensions of $u(t)$ and $y(t)$ to be $4$, although these can be adjusted based on practical needs.}. The relative weight matrices for the LQG performance are set as $Q = I$ and $R = 0.1I$, respectively. The process and measurement noises are supposed to be $\Pi_{w} = 0.01I$ and $\Pi_v = 0.1I$. In the following, we will conduct simulations with the above model setup for all three configuration scenarios, i.e., sparsity promoting configuration, low-rank promoting configuration, and structure-constrained configuration. 

Throughout this simulation, the ADMM penalty parameter is chosen as $\rho = 1$, the parameters of Algorithm 1 are chosen as $\alpha = \beta = 0.5$, and the tolerances are chosen as $\epsilon^{\text{pri}}=\epsilon^{\text{{dual}}}=\epsilon^{\text{{main}}}=1e-5$. We first choose the relative weight for the configuration cost as $\gamma = 0.01$ and initialize the sensor and actuator matrices randomly as follows:
\begin{align*}
     B_0 = \begin{bmatrix}
   0.6281 & 0.4653 & 1.3792 & 0.7538 \\
0.7641 & -1.0122 & -0.4788 & -0.7347 \\
-0.6719 & -1.1430 & 0.2558 & 0.0179 \\
0.5391 & -0.0049 & -0.8286 & 0.5972
    \end{bmatrix},
\end{align*}
and 
\begin{align*}
     C_0 = \begin{bmatrix}
1.1727 & -1.0521 & 0.9429 & -0.9102 \\
1.3779 & -0.7275 & -0.7694 & -0.7467 \\
0.1416 & 1.0222 & 1.2878 & 0.3481 \\
-1.0841 & -0.7372 & -0.4500 & -0.0801
    \end{bmatrix}.
\end{align*}
% By solving the algebraic Riccati equations in \eqref{care1}--\eqref{care2}, the initial LQG performance is obtained as $J_{\text{LQG}}^0 =  6.2688$.

\subsection{Simulation I: Sparsity Promoting Configuration}
\begin{figure}[t]
	\centering	\includegraphics[width=0.39\textwidth]{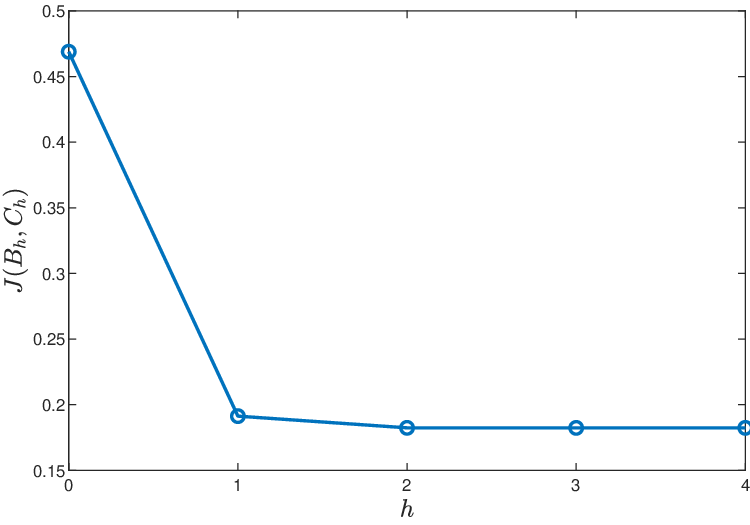}
	\caption{Convergence process of Algorithm 2 in Problem S1.
	\vspace{-15pt}}
\end{figure}

In this part, we consider the sparsity promoting configuration scenario and employ Algorithm 2 for numerical computation.
As shown in Fig. 1, Algorithm 2 converges after 4 iterations, reducing the overall cost from $0.4689$ to $0.1823$, and the following sensor and actuator matrices are returned:
\begin{align*}
     B^* = \begin{bmatrix}
0 & 0 & 2.3997 & 0 \\
0 & -0.0744 & 0 & 0 \\
0 & 0 & 0 & 0 \\
0 & -0.6850 & 0 & 0
    \end{bmatrix},
\end{align*}
and 
\begin{align*}
     C^* = \begin{bmatrix}
5.4692 & 0 & 0 & 0 \\
0 & 0 & 0 & 0 \\
0 & 0 & 0 & 2.3136 \\
0 & 0 & 0 & 0
    \end{bmatrix},
\end{align*}
which exhibit much sparser structures. The optimal controller gain matrix and observer gain matrix are given by 
\begin{align*}
     K^* = \begin{bmatrix}
0 & 0 & 0 & 0 \\
-0.3158 & 2.8312 & 0.6573 & 3.2382 \\
-4.0133 & -0.0820 & -1.8834 & 1.1153 \\
0 & 0 & 0 & 0
    \end{bmatrix},
\end{align*}
and 
\begin{align*}
     L^* = \begin{bmatrix}
-0.9575 & 0 & -0.0310 & 0 \\
0.0550  & 0 & -0.0619 & 0 \\
-0.1964 & 0 & -0.2908 & 0 \\
-0.0732 & 0 & -0.3109 & 0
    \end{bmatrix}.
\end{align*}
It is easy to verify that both $A+B^*K^*$ and $A+L^*C^*$ are Hurwitz matrices. Therefore, $(A,B)$ is stabilizable and $(A,C)$ is detectable.
The associated LQG performance and configuration cost are, respectively, obtained as follows:
\begin{align*}
    J_{\text{LQG}}^* =  0.0729~~\text{and}~~\tilde{J}_{\text{SAC}}^* = 10.9419~~({J}_{\text{SAC}}^* = 5).
\end{align*}
To verify the stationary solution, by Corollary 1, we have that
\begin{align*}
   P(G_1+G_2)PB^*R^{-1} =  \begin{bmatrix}
        0 & -0.0004 & \underline{0.0050} & 0 \\
0 & \underline{-0.0050} & 0.0021 & 0 \\
0 & -0.0023 & 0.0038 & 0 \\
0 & \underline{-0.0050} & 0.0013 & 0
    \end{bmatrix}.
\end{align*}
where all elements have magnitudes less than or equal to $\gamma/2 = 0.005$, with equality achieved at the nonzeros in $B^*$ (as underlined). Similarly, we have $\Pi_v^{-1}CX(H_1+H_2)X = $
\begin{align*}
    \begin{bmatrix}
\underline{0.0050} & -0.0004 & 0.0010 & 0.0002 \\
0 & 0 & 0 & 0 \\
0.0001 & 0.0014 & 0.0048 & \underline{0.0050} \\
0 & 0 & 0 & 0
\end{bmatrix},
\end{align*}
where all elements have magnitudes less than or equal to $\gamma/2 = 0.005$, with equality achieved at the nonzeros in $C^*$. Therefore we conclude that $(B^*,C^*)$ is a stationary point.

Furthermore, we also design the sensor and actuator matrices using different relative weights for the configuration cost. The numerical results are summarized in Fig. 2, which illustrates a tradeoff between the LQG performance $J_{\text{LQG}}$ and the configuration cost $J_{\text{SAC}}$. Meanwhile, note that $\tilde{J}_{\text{SAC}}$ and $J_{\text{SAC}}$ exhibit similar changing trends in Fig. 3, which verifies the effectiveness of using the $\ell_1$ norm to promote sparsity.

\begin{figure}[t]
	\centering	\includegraphics[width=0.39\textwidth]{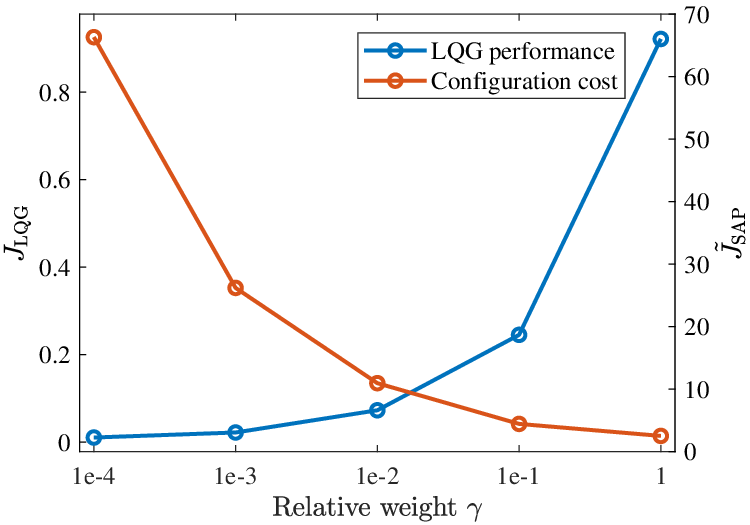}
	\caption{The LQG performance $J_{\rm LQG}$ versus the configuration cost $\tilde{J}_{\rm SAC}$ under different relative weights $\gamma$.
	\vspace{-5pt}}
\end{figure}
\begin{figure}[t]
	\centering	\includegraphics[width=0.39\textwidth]{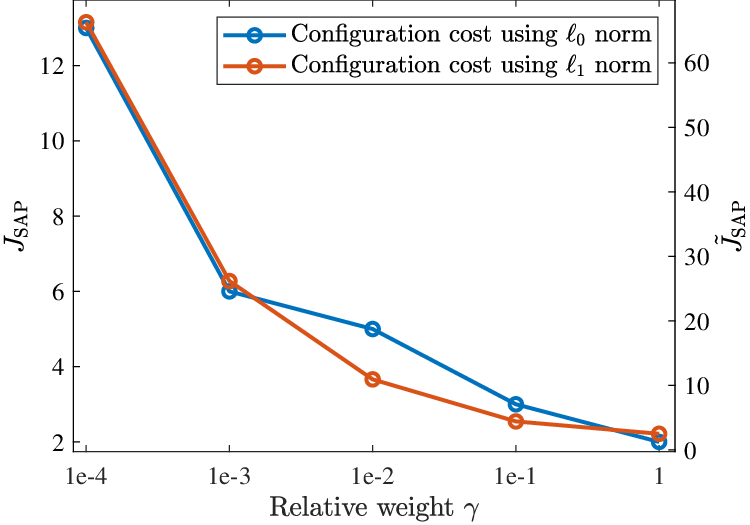}
	\caption{The configuration costs defined using the $\ell_0$ norm and $\ell_1$ norm under different relative weights $\gamma$.
	\vspace{-15pt}}
\end{figure}

\subsection{Simulation II: Low-Rank Promoting Configuration}
\begin{figure}[t]
	\centering	\includegraphics[width=0.38\textwidth]{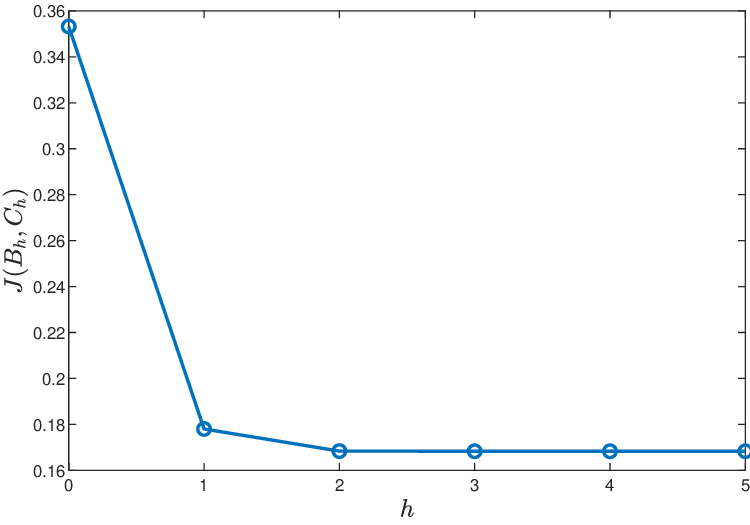}
	\caption{Convergence process of Algorithm 2 in Problem S2.
	\vspace{-15pt}}
\end{figure}
In this part, we consider the low-rank promoting configuration scenario and use Algorithm 2 for numerical computation.
As shown in Fig. 4, Algorithm 2 converges after 5 iterations, reducing the overall cost from $0.3532 $ to $0.1683$, and the following sensor and actuator matrices are returned:
\begin{align*}
     B^* = \begin{bmatrix}
 0.4133 & -0.6914 & 1.6144 & 0.5315 \\
0.2048 & -0.3033 & -0.1795 & 0.0359 \\
0.2888 & -0.4633 & 0.6323 & 0.2563 \\
0.1973 & -0.2807 & -0.4621 & -0.0325
    \end{bmatrix},
\end{align*}
and 
\begin{align*}
     C^* = \begin{bmatrix}
3.2953 & -0.2480 & 0.4490 & -0.0916 \\
2.9990 & -0.4752 & -0.3988 & -0.9395 \\
0.8471 & 0.2633 & 1.1735 & 1.0983 \\
-2.7723 & 0.0556 & -0.8727 & -0.4477
    \end{bmatrix},
\end{align*}
which exhibit low-rank structures since $\rank(B^*)=2$ and $\rank(C^*)=2$.
By solving \eqref{care1}--\eqref{care2}, the optimal controller gain matrix and observer gain matrix are given by 
\begin{align*}
     K^* = \begin{bmatrix}
 -0.8053 & -1.4105 & -0.8841 & -1.1842 \\
1.3631 & 2.0820 & 1.3975 & 1.6890 \\
-3.5413 & 1.4138 & -1.4179 & 2.6598 \\
-1.1274 & -0.2061 & -0.6642 & 0.1689
    \end{bmatrix},
\end{align*}
and 
\begin{align*}
     L^* = \begin{bmatrix}
 -0.5779 & -0.5178 & -0.1592 & 0.4912 \\
0.0381 & 0.0767 & -0.0454 & -0.0062 \\
-0.0926 & 0.0935 & -0.2567 & 0.1868 \\
-0.0164 & 0.1691 & -0.2452 & 0.1265
    \end{bmatrix}.
\end{align*}
It is easy to verify that both $A+B^*K^*$ and $A+L^*C^*$ are Hurwitz matrices by checking their eigenvalues. Therefore, we have that $(A,B)$ is stabilizable and $(A,C)$ is detectable.
The associated LQG performance and configuration cost are, respectively, obtained as follows:
\begin{align*}
    J_{\text{LQG}}^* =0.0657~~\text{and}~~\tilde{J}_{\text{SAC}}^* =   10.2619 ~~({J}_{\text{SAC}}^* =   4).
\end{align*}
It is worth noticing that, by Lemma 2, we can further obtain the following rank factorization:
\begin{align*}
    B^* = B_1B_2~~ \text{  and  }~~ C^* = C_1C_2
\end{align*}
where
\begin{align*}
    &B_1 = \begin{bmatrix}
        -1.3041 & 0.0279 \\
-0.0122 & -0.4925 \\
-0.5904 & -0.2369 \\
0.1754 & -0.6241
    \end{bmatrix},\\
&B_2=\begin{bmatrix}
    -0.3256 & 0.5431 & -1.2295 & -0.4089 \\
-0.4077 & 0.6024 & 0.3949 & -0.0628
\end{bmatrix},    
\end{align*}
and
\begin{align*}
    &C_1 = \begin{bmatrix}
   -1.4383 & -0.0016 \\
-1.2799 & -0.8278 \\
-0.4078 & 1.0825 \\
1.2278 & -0.5052
    \end{bmatrix},\\
&C_2=\begin{bmatrix}
 -2.2911 & 0.1721 & -0.3133 & 0.0626 \\
-0.0806 & 0.3081 & 0.9660 & 1.0381
\end{bmatrix}.   
\end{align*}
Therefore, we can use \( B_1 \) and \( C_2 \) as the actuator and sensor matrices, with $\tilde{u}(t)=B_2u(t)$ and $\tilde{y}(t) = C_2x(t)$ as intermediate signals. In this way, the number of installed actuators as well as sensors can be reduced from 4 to 2.
\begin{figure}[t]
	\centering	\includegraphics[width=0.39\textwidth]{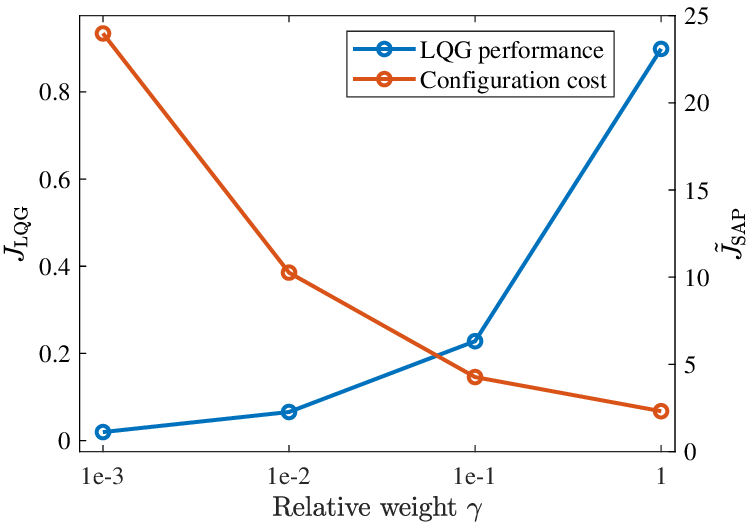}
	\caption{The LQG performance $J_{\rm LQG}$ versus the configuration cost $\tilde{J}_{\rm SAC}$ under different relative weights $\gamma$.
	\vspace{-5pt}}
\end{figure}
\begin{figure}[t]
	\centering	\includegraphics[width=0.39\textwidth]{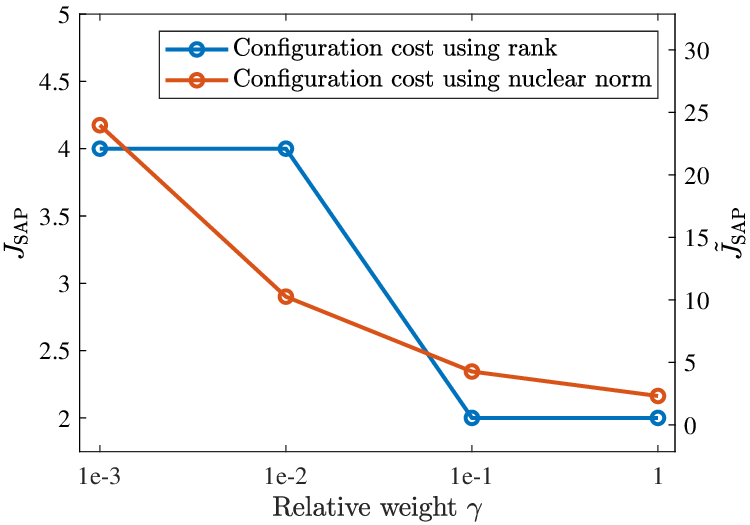}
	\caption{The configuration costs defined using the nuclear norm and matrix rank under different relative weights $\gamma$.
	\vspace{-15pt}}
\end{figure}

To verify the stationary solution, by Corollary 2, we first compute that $ P(G_1+G_2)PB^*R^{-1} = $
\begin{align*}
 \begin{bmatrix}
0.0009 & -0.0016 & 0.0039 & 0.0013 \\
0.0015 & -0.0022 & -0.0014 & 0.0002 \\
0.0012 & -0.0018 & 0.0010 & 0.0007 \\
0.0017 & -0.0025 & -0.0023 & 0.0001
    \end{bmatrix}.
\end{align*}
Then we do the singular value decomposition:
\begin{align*}
B^* &= U_B\Sigma_BV_B^{\T}= \\
    &\begin{bmatrix}
-0.9042 & 0.0337 & 0 & 0 \\
-0.0085 & -0.5933 & 0 & 0 \\
-0.4093 & -0.2854 & 0 & 0 \\
0.1216 & -0.7519 & 0 & 0
\end{bmatrix} \begin{bmatrix}
2.0800 & 0 & 0 & 0 \\
0 & 0.6890 & 0 & 0 \\
0 & 0 & 0 & 0 \\
0 & 0 & 0 & 0
\end{bmatrix}\\
&~~~~~~~~~~~~~\times \begin{bmatrix}
-0.2258 & 0.3766 & -0.8525 & -0.2835 \\
-0.4912 & 0.7257 & 0.4758 & -0.0756 \\
0 & 0 & 0 & 0 \\
0 & 0 & 0 & 0
\end{bmatrix}.
\end{align*}
It is easy to verify that 
\begin{align*}
     \| 2 P(G_1+G_2)PB^*R^{-1} -\gamma U_BV_B^{\T} \|_{2}&=1.3975e-04 <\gamma,\\
     2 U_B^{\T}P(G_1+G_2)PB^*R^{-1} &\approx\gamma V_B^{\T},\\
     2 P(G_1+G_2)PB^*R^{-1}V_B &\approx \gamma U_B.
\end{align*}
Therefore, the conditions in \eqref{c21m} are satisfied. The conditions for $C^*$ can be verified similarly and are omitted here. Finally, we can conclude that $(B^*,C^*)$ is a stationary point.
        
Moreover, we also design the sensor and actuator matrices with different relative weights for the configuration cost. The numerical results, as illustrated in Fig. 5, reveal a tradeoff between the LQG performance $J_{\text{LQG}}$ and the configuration cost $J_{\text{SAC}}$. Furthermore, it can be observed from Fig. 6 that the numerical values of $\tilde{J}_{\text{SAC}}$ and $J_{\text{SAC}}$ exhibit similar changing trends, which verifies the effectiveness of using the nuclear norm to promote low rank solutions.

\subsection{Simulation III: Structure-Constrained Configuration}
\begin{figure}[t]
	\centering	\includegraphics[width=0.38\textwidth]{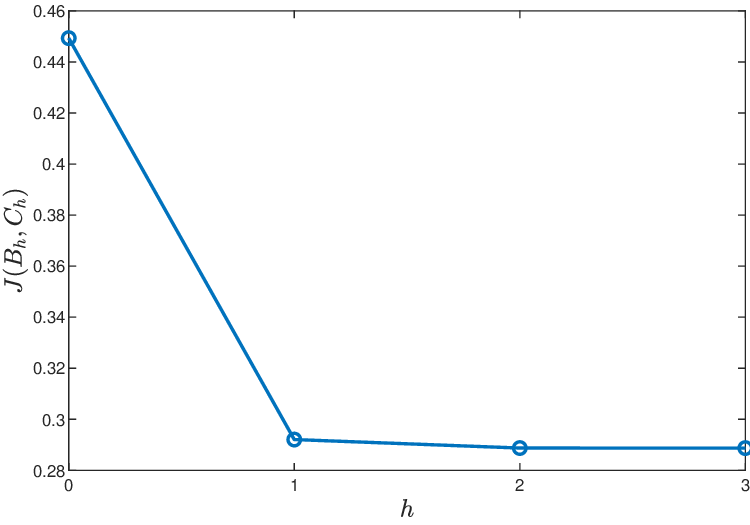}
	\caption{Convergence process of Algorithm 2 in Problem S3.
	\vspace{-15pt}}
\end{figure}
In this part, we consider the structure-constrained configuration scenario, where the following structures are imposed:
\begin{align}\label{struct}
     S_B = \begin{bmatrix}
1 & 0 & 0 & 1 \\
0 & 1 & 0 & 0 \\
0 & 0 & 1 & 0 \\
1 & 0 & 0 & 0
\end{bmatrix} ~~\text{and}~~S_C=\begin{bmatrix}
0 & 1 & 0 & 0 \\
0 & 0 & 0 & 1 \\
1 & 0 & 0 & 0 \\
0 & 0 & 1 & 0
\end{bmatrix}.
\end{align}
The initial values $B^0$ and $C^0$ are obviously infeasible configuration. Therefore, we use Algorithm 2 for structural design. 
As shown in Fig. 7, Algorithm 2 converges after 3 iterations, reducing the overall cost from $0.4493$ to $0.2887$, and the following sensor and actuator matrices are returned:
\begin{align*}
     B^* = \begin{bmatrix}
  \underline{0} & -1.5731 & 0.1713 &  \underline{0} \\
0.0000 &  \underline{0} & -0.1819 & -0.3388 \\
0.0000 & -0.7150 &  \underline{0} & -0.1682 \\
 \underline{0} & 0.2883 & -0.2529 & -0.4041
    \end{bmatrix},
\end{align*}
and 
\begin{align*}
     C^* = \begin{bmatrix}
 1.2307 &  \underline{0} & 0.4348 & 0.2890 \\
1.8847 & -0.1439 & 0.2450 &  \underline{0} \\
 \underline{0} & 0.2174 & 0.8294 & 0.8628 \\
-1.9356 & 0.2124 &  \underline{0} & 0.2512
    \end{bmatrix},
\end{align*}
which satisfy the prescribed structures in \eqref{struct}. The optimal controller gain matrix and observer gain matrix are given by 
\begin{align*}
     K^* = \begin{bmatrix}
 -0.0000 & -0.0000 & -0.0000 & -0.0000 \\
4.1789 & 0.1273 & 2.3008 & -1.4346 \\
-0.4672 & 1.4021 & 0.1883 & 1.5796 \\
0.0040 & 2.6360 & 0.8440 & 2.6343
    \end{bmatrix},
\end{align*}
and 
\begin{align*}
     L^* = \begin{bmatrix}
  -0.6160 & -0.9367 & -0.0159 & 0.9570 \\
0.0261 & 0.0726 & -0.0627 & -0.0932 \\
-0.1757 & -0.1178 & -0.2945 & 0.0335 \\
-0.1096 & -0.0128 & -0.3012 & -0.0763
    \end{bmatrix}.
\end{align*}
It is easy to verify that both $A+B^*K^*$ and $A+L^*C^*$ are Hurwitz matrices. Therefore, $(A,B)$ is stabilizable and $(A,C)$ is detectable.
The associated LQG performance and configuration cost are, respectively, obtained as follows:
\begin{align*}
    J_{\text{LQG}}^* = 0.1461~~\text{and}~~J_{\text{SAC}}^* = 14.2558.
\end{align*}
To verify the stationary solution, by Corollary 3, we compute that $\|[P(G_1+G_2)PB^*R^{-1} -\gamma B^*]\circ \bar{S}_B \|_{\text{F}} = 5.1999e-5 \approx 0$ and $\|[\Pi_v^{-1}C^*X(H_1+H_2)X-\gamma C^*] \circ \bar{S}_C \|_{\text{F}}=4.8495e-5\approx 0$. Thus we can conclude that $(B^*,C^*)$ is a stationary point.
We also design the sensor and actuator matrices by using different relative weights for the configuration cost. The numerical results are shown in Fig. 8, revealing a tradeoff between the LQG performance $J_{\text{LQG}}$ and the configuration cost $J_{\text{SAC}}$. 
\begin{figure}[t]
	\centering	\includegraphics[width=0.39\textwidth]{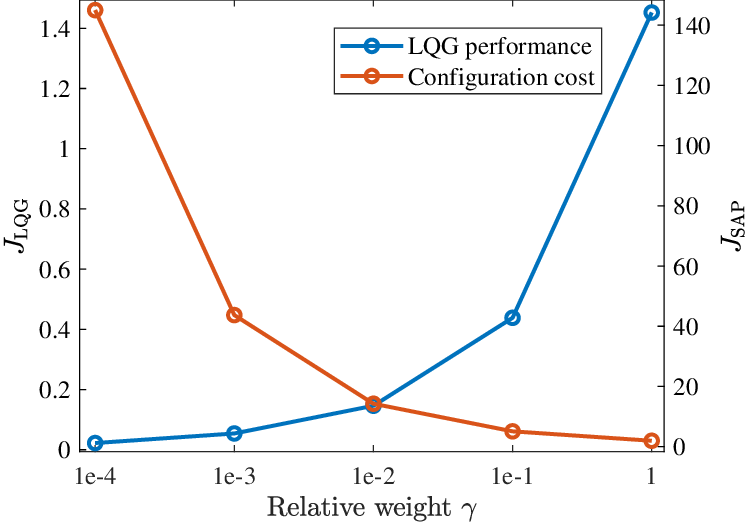}
	\caption{The LQG performance $J_{\rm LQG}$ versus the configuration cost ${J}_{\rm SAC}$ under different relative weights $\gamma$.
	\vspace{-15pt}}
\end{figure}
\section{Conclusions}
In this paper, we studied the joint configuration of sensors and actuators under the LQG performance. Unlike existing research which focuses on selecting or placing sensors and actuators from predefined candidates, we introduced a new formulation in which sensors and actuators are designed subject to general-form constraints and costs. We derived the analytical gradients of the LQG performance with respect to the sensor and actuator matrices using algebraic Riccati equations and established the first-order optimality conditions based on KKT analysis. Furthermore, we developed a tractable ADMM-based alternating optimization framework, which ensures computational efficiency and adaptability to various configuration constraints and costs.
We also investigated three specific configuration scenarios: sparsity promoting, low-rank promoting and structure-constrained configurations, and provide tailored computational schemes. The effectiveness of our results were well verified by numerical simulations. Future research include extending this framework to nonlinear systems and exploring real-time adaptation mechanisms in dynamic environments.

\bibliographystyle{IEEEtran}
\bibliography{main}

\end{document}